\documentclass[showpacs,amssymb,preprint,preprintnumbers]{revtex4}

\usepackage{amsmath}

\begin{document}

\title{Wightman function and stochastic gravity noise kernel in impulsive plane wave spacetimes}
\author{Hing-Tong Cho}
\email[Email: ]{htcho@mail.tku.edu.tw}
\affiliation{Department of Physics, Tamkang University, Tamsui, New Taipei City, TAIWAN}

\begin{abstract}
In this paper we study quantum field theory in impulsive plane wave spacetimes. We first analyze the geodesics and the formation of conjugate planes in these spacetimes. The behaviors of the world function and the van Vleck determinant near conjugate plane are also considered. For the quantum field, we work out the mode functions, their Bogoliubov transformations, and the construction of the Wightman functions. By examining the Wightman function near and on the conjugate plane, we show how the twofold and fourfold singularity structure of the Wigthman function arise when crossing this plane. Lastly, we come to the stochastic gravity noise kernel which is also the correlation function of the stress energy tensor of the quantum field. Its explicit form is given in terms of the world function and the van Vleck determinant. We investigate its limits for small and large geodesic distances. The leading divergent term of the noise kernel on the conjugate plane are expressed in terms of derivatives of delta functions. Similar to that of the Wightman functions, we also examine how the singularity structure of the noise kernel near the lightcone changes when crossing the conjugate plane.
\end{abstract}

\date{\today}
\maketitle

\section{Introduction}
Our interest in plane wave spacetimes originates from the Penrose limiting procedure \cite{Pen76} in which a general spacetime can be transformed into a plane wave one. It basically describes the spacetime near a null geodesic and one hopes to capture the essential physics in this region through this limiting procedure. For example, Shore and collaborators \cite{HolSho08,HSS09} have utilized this limit to prove that photon propagation in QED including the fermion quantum effect is causal in a general curved spacetime despite the appearance of superluminal low-frequency phase velocities \cite{DruHat80}. More recently, this method has been used to consider the memory effect of geodesic congruences in gravitational shockwave spacetimes \cite{Shore18}.

Another interest example in this approach is about the global Green function of a massless scalar field propagating in curved spacetimes. A prominent case would be that of a black hole spacetime. Due to the presence of caustic points around a black hole where the neighboring null geodesics focus, it has been discovered that the leading singularity of the retarded Green function near the lightcone changes after passing through such a caustic point \cite{CDOW09}. In fact, this change possesses a fourfold structure, $\delta(\sigma)\rightarrow 1/\sigma\rightarrow-\delta(\sigma)\rightarrow-1/\sigma\rightarrow\delta(\sigma)\rightarrow\cdots$ \cite{BusCas18}. This structure is important in the understanding of wave propagation in general curved spacetimes. In \cite{HarDri12}, it has been shown that these fourfold and sometimes twofold structures emerge for the Green function in plane wave spacetimes. Through the Penrose limit, they argue that these patterns are also valid for general curved spacetimes.

Impulsive plane waves have $\delta$ function profiles. Before the arrival and after the passing of the wave, the spacetimes are Minkowski. The spacetime is thus simple enough that exact solutions for the geodesics as well as the mode functions of the quantum scalar field can be obtained \cite{Gibbons75,GarVer91}. The authors in \cite{ZDH18} have taken advantage of this fact to consider the memory effect of impulsive gravitational waves. They have found a velocity memory in which particles initially at rest moves apart or towards each other after the passage of the wave. This velocity memory actually is the reason why neighboring null geodesics focus to form caustic points. These caustic points are located on what are called the conjugate planes. The geodesic distance diverges near these planes \cite{HarDri12} so that the usual construction of correlation functions like the Wightman function may not apply. Here, with the simple setting of impulsive plane wave spacetimes, we hope to study the Wightman function on conjugate planes and their singular structures near the lightcone across these planes in an analytical way.

In the semiclassical gravity theory, the interaction between the quantum field and the classical spacetime is governed by the Einstein equation with the source described by the expectation value of the stress energy tensor of the field \cite{BirDav82}. This can be thought of as a mean field theory. To account for fluctuation and correlation effects, Hu and Verdaguer have devised an open quantum system approach \cite{FeyVer63} called the stochastic gravity theory \cite{HuVer08,HuVer20}. In stochastic gravity, the environment is the quantum field and the system the classical spacetime. The effects of the quantum field, in addition to the expectation value of the stress energy tensor, are also represented by the dissipation and the noise kernels. The semiclassical Einstein equation is replaced by the Einstein-Langevin equation with a stochastic tensor force. The correlation function of this force is the noise kernel which is just the correlator of the quantum field stress tensor \cite{PhiHu01}.
To explore the physics near a null geodesic in a general curved spacetime, for example, the quantum energy inequalities \cite{ForRom95}, one would like to apply the stochastic gravity approach to plane wave spacetimes under the Penrose limit. As a first step in this consideration, we shall in the following take a close look at the behaviors of the noise kernel near and on the conjugate planes in these impulsive plane wave spacetimes. 

In the next section we solve the geodesic equations in impulsive plane wave spacetimes. Two different cases are considered, the degenerate one corresponding to a gravitational plane wave while the nondegenerate one an electromagnetic wave. We show that the velocity memory effects occur in both cases, and 
these effects focus the null geodesics to form conjugate planes. The world function and the van Vleck determinant which are needed to construct the correlation functions like the Wightman function in these spacetimes are then defined. Finally, we explore the properties of these bitensors especially when they approach the conjugate plane. 

In Sec.~III, we turn our attention to quantum field theory in these spacetimes. The main goal is to work out the Wightman function of a scalar field and to study its properties \cite{BogShi80}. To do that we first obtain the in- and out-mode functions by solving the corresponding Klein-Gordon equation \cite{Gibbons75,Klimcik88}. Then, we look at their focusing behavior near the conjugate plane as well as their Bogoliubov transformations. With these mode functions, we contruct the correlation functions, in particular, the Wightman function. Their explicit form on the conjugate plane are given. We also show how their singularity structure near the lightcone on both sides of this plane emerge.

We consider the noise kernel in stochastic gravity in Sec.~IV. Using the method of point-separation, we give the explicit expression for this noise kernel which is also the correlation function of the stress energy tensor of the quantum field \cite{PhiHu01}. This consideration will be useful when we apply the stochastic gravity theory to plane wave spacetimes. We also examine the properties of the noise kernel on and near the conjugate plane. Lastly, the conclusions and discussions are presented in Sec.~V.

\section{Impulsive plane wave spacetimes: Geodesics, memory and conjugate planes}
In this section we shall examine the geodesics in the impulsive plane wave spacetime and the corresponding memory effects that these geodesics entail. The spacetime can be represented by the metric \cite{Blau11}
\begin{eqnarray}
ds^{2}=-2dudv+\sum_{a,b=1}^{2}H_{ab}\,\delta(u)\,x^{a}x^{b}du^{2}+\sum_{a=1}^{2}dx^{a}dx_{a},
\end{eqnarray}
in which the shockwave is located at $u=0$. Without loss of generality, one can set the $2\times 2$ matrix
\begin{eqnarray}
H_{ab}=\left(
\begin{array}{cc}
\lambda_{1} & 0 \\ 0 & \lambda_{2}
\end{array}
\right).
\end{eqnarray}
Hence, the metric can be rewritten as
\begin{eqnarray}
ds^{2}=-2dudv+\delta(u)\,f(\vec{x})\,du^{2}+\sum_{a=1}^{2}dx^{a}dx_{a},\label{planemetric}
\end{eqnarray}
where $f(\vec{x})=\sum_{a=1,2}\lambda_{a}(x^{a})^{2}$. The only nonvanishing Ricci tensor component $R_{uu}$ at the wavefront is proportional to the trace of $H_{ab}$ \cite{GarVer91}. If $H_{ab}$ is traceless, that is, ${H_{a}}^{a}=0$, it corresponds to a Ricci flat spacetime. Then, the wave can be considered as a pure gravitational one. A typical case we shall explore in some detail in this paper is with $\lambda_{1}=-\lambda=-\lambda_{2}$. Due to the weak energy condition, we require $\lambda\geq 0$. Another case we shall also be interested in is with $\lambda_{1}=\lambda_{2}=-\lambda$.  Here the Weyl tensor vanishes and the wave corresponds to that of a pure electromagnetic one.

\subsection{Geodesics and memory}
The set of geodesic equations corresponding to the metric in Eq.~(\ref{planemetric}) is
\begin{eqnarray}
&&\frac{d^{2}u}{d\eta^{2}}=0\nonumber\\
&&\frac{d^{2}v}{d\eta^{2}}-\frac{1}{2}\delta'(u)\sum_{a=1,2}\lambda_{a}(x^{a})^{2}\left(\frac{du}{d\eta}\right)^{2}-2\delta(u)\sum_{a=1,2}\lambda_{a}x^{a}\frac{dx^{a}}{d\eta}\frac{du}{d\eta}=0\nonumber\\
&&\frac{d^{2}x^{a}}{d\eta^{2}}-\delta(u)\lambda_{a}x^{a}\left(\frac{du}{d\eta}\right)^{2}=0
\end{eqnarray}
where there is no sum over $a$ in the last equation, and $\eta$ is an affine parameter. The $u$-equation can be solved readily to give
\begin{eqnarray}
u(\eta)=\left(\frac{du}{d\eta}\right)_{\eta=0}\eta+u(0).
\end{eqnarray}
From this solution, one can see that it is possible to treat $u$ as the affine parameter for null and timelike geodesics. If we take $\eta=u$, the $x^{a}$-equation simplifies to
\begin{eqnarray}
\ddot{x^{a}}=\delta(u)\lambda_{a}x^{a}\label{xequation}
\end{eqnarray}
where the overdot represents derivative with respect to $u$. For $u\neq 0$, the solutions are straight lines. For $u<0$, take the boundary conditions, $\dot{x}^{a}|_{u=0}=\dot{x}^{a}_{0}$ and $x^{a}|_{u=0}=x_{0}^{a}$. The solution is $x^{a}_{<}=\dot{x}^{a}_{0}u+x_{0}^{a}$. To obtain the solution for $u>0$, we note that Eq.~(\ref{xequation}) requires the solutions for $u<0$ and $u>0$ to be continuous and their derivatives to differ by 
\begin{eqnarray}
\dot{x}_{>}^{a}|_{u=0}-\dot{x}_{<}^{a}|_{u=0}=\lambda_{a}x_{0}^{a}\label{velmemory}
\end{eqnarray}
Therefore, 
\begin{eqnarray}
x_{>}^{a}=(\lambda_{a}x_{0}^{a}+\dot{x}_{0}^{a})u+x_{0}^{a}
\end{eqnarray}
Combining with the $x_{<}^{a}$ result, the solution to the $x^{a}$-equation can be expressed as
\begin{eqnarray}
x^{a}=u\,\theta(u)\lambda_{a}x_{0}^{a}+\dot{x}_{0}^{a}u+x_{0}^{a},\label{xsolution}
\end{eqnarray}
where $\theta(u)$ is the step function.

Next, we consider the $v$-equation, and we note that, using the geodesic equations, the quantity
\begin{eqnarray}
\xi\equiv\left(\frac{ds}{du}\right)^{2}=-2\frac{dv}{du}+\delta(u)\sum_{a=1,2}\lambda_{a}(x^{a})^{2}+\sum_{a=1,2}\frac{dx_{a}}{du}\frac{dx^{a}}{du}\label{defxi}
\end{eqnarray}
is a constant of motion with $\xi<0$ and $\xi=0$ for timelike and null geodesics, respectively. The solution to this equation, with $v(0^{-})=v_{0}$, is given by
\begin{eqnarray}
v=v_{0}-\frac{1}{2}\xi u+\sum_{a=1,2}\left[\frac{1}{2}(\dot{x}_{0}^{a})^{2}u+\frac{1}{2}\theta(u)\lambda_{a}(x_{0}^{a})^{2}+u\,\theta(u)\left(\frac{1}{2}\lambda_{a}^{2}(x_{0}^{a})^{2}+\lambda_{a}x_{0}^{a}\dot{x}_{0}^{a}\right)\right].\label{vsolution}
\end{eqnarray}
$v(u)$ is discontinuous across the shock at $u=0$,
\begin{eqnarray}
v(0^{+})-v(0^{-})=\frac{1}{2}\sum_{a=1,2}\lambda_{a}(x_{0}^{a})^{2},\label{vdiscon}
\end{eqnarray}
as required by the presence of the delta function term in Eq.~(\ref{defxi}).

The discontinuity in Eq.~(\ref{velmemory}) has been interpreted as what is called the memory effect \cite{ZDH18}. After the passage of the shockwave, the velocity of the particle changes abruptly from $\dot{x}_{<}^{a}|_{u=0}$ to $\dot{x}_{>}^{a}|_{u=0}$. Since this ``velocity memory effect'' is proportional to $x_{0}^{a}$, neighboring parallel geodesics will therefore diverge ($\lambda_{a}>0$) or converge ($\lambda_{a}<0$) after interacting with the shock. 

\subsection{Focusing of geodesics and conjugate planes}
Due to the discontinuity of the particle velocities discussed above, the phenomenon of geodesic focusing or caustics will occur \cite{GarVer91}. To analyze this more concretely, we consider the degenerate case in which $\lambda_{1}=\lambda_{2}=-\lambda$. In addition, we take $\dot{x}_{0}^{a}=0$, that is, for perpendicular incidence, the geodesics are given by Eqs.~(\ref{xsolution}) and (\ref{vsolution}),
\begin{eqnarray}
x^{a}&=&(1-u\,\lambda)\, x_{0}^{a},\nonumber\\
v&=&v_{0}-\frac{1}{2}\,\xi\, u-\frac{1}{2}\lambda(1-u\lambda)\sum_{a=1,2}(x_{0}^{a})^{2}.
\end{eqnarray}
for $u>0$. At $u=1/\lambda$, irrespective of the value of $x_{0}^{a}$, these geodesics will all focus to a point with $x^{a}=0$ and $v=v_{0}-\xi/2\lambda$.

This focusing of geodesics will also happen in the nondegenerate case with $\lambda_{1}=-\lambda=-\lambda_{2}$. Then, we have for $u>0$,
\begin{eqnarray}
x^{1}=x^{1}_{0}(1-u\lambda)\ \ \ ;\ \ \ x^{2}=x^{2}_{0}(1+u\lambda).
\end{eqnarray}
At the focal plane $u=1/\lambda$, $x^{1}=0$ and $x^{2}=2x_{0}^{2}$. The geodesic congruence will map to a line in the $x^{1}$-$x^{2}$ plane, rather than a point in the previous degenerate case. The corresponding $v$ coordinate is
\begin{eqnarray}
v=v_{0}-\frac{1}{2}\,\xi\,u-\frac{1}{2}\lambda[(x_{0}^{1})^{2}(1-u\lambda)-(x_{0}^{2})^{2}(1+u\lambda)].
\end{eqnarray}
At $u=1/\lambda$, 
\begin{eqnarray}
v=v_{0}-\frac{\xi}{2\lambda}+\lambda(x_{0}^{2})^{2}
\Rightarrow v=\frac{\lambda}{4}(x^{2})^{2}+\left(v_{0}-\frac{\xi}{2\lambda}\right).
\end{eqnarray}
With fixed $v_{0}$, this is a parabola in the $v$-$x^{2}$ plane.

Focusing of geodesics, especially for the null geodesics, leads to the existence of conjugate planes in this impulsive plane wave spacetime \cite{HarDri12}. To define these conjugate planes, it is more convenient to describe the geodesics in terms of their starting position $(v',{x'}^{a})$ and velocity $(\dot{v}',\dot{x}'$$^{a})$. To do so, we express $x_{0}^{a}$ and $\dot{x}^{a}_{0}$ in terms of $(v',{x'}^{a})$ and $(\dot{v}',\dot{x}'$$^{a})$ from Eq. (\ref{xsolution}) with $u'<0$,
\begin{eqnarray}
x_{0}^{a}=-\dot{x}'\ \!\!^{a}u'+x'\ \!\!^{a}\ \ \ ;\ \ \ \dot{x}_{0}^{a}=\dot{x}'\ \!\!^{a}.
\end{eqnarray}
The geodesic equation in Eq.~(\ref{xsolution}) then becomes
\begin{eqnarray}
x^{a}=[1+u\theta(u)\lambda_{a}]\,x'\ \!\!^{a}+[(u-u')-uu'\theta(u)\lambda_{a}]\,\dot{x}'\ \!\!^{a}.\label{xsolution1}
\end{eqnarray}
Similarly, from Eq.~(\ref{vsolution}) and replacing $\dot{x}_{0}^{a}$ by $\dot{x}'\ \!\!^{a}$, we have
\begin{eqnarray}
v_{0}=v'+\frac{1}{2}\xi u'-\frac{1}{2}\sum_{a=1,2}(\dot{x}'\ \!\!^{a})^{2}u'
\end{eqnarray}
In terms of $v'$, the geodesic equation for $v$ is
\begin{eqnarray}
v&=&v'-\frac{1}{2}\xi(u-u')+\sum_{a=1,2}\left\{\frac{1}{2}\lambda_{a}\theta(u)(1+u\lambda_{a}){x'}^{a}{x'}^{a}+\lambda_{a}\theta(u)(u-u'-\lambda_{a}uu'){x'}^{a}\dot{x}'\ \!\!^{a}\right.\nonumber\\
&&\hskip 120pt \left.+\left[\frac{1}{2}(u-u')-\frac{1}{2}\lambda_{a}u'\theta(u)(2u-u'-uu'\lambda_{a})\right]\dot{x}'\ \!\!^{a}\dot{x}'\ \!\!^{a}
\right\}\label{vsolution1}
\end{eqnarray}

Conjugate planes occur when geodesics focus to a point or a line. For example, in the degenerate case, $\lambda_{1}=\lambda_{2}=-\lambda$, the $x^{a}$ coordinates in Eq.~(\ref{xsolution1}) depend on ${x'}^{a}$ and $\dot{x}'$$^{a}$ as
\begin{eqnarray}
x^{a}=(1-u\lambda)\,x'\ \!\!^{a}+(u-u'+uu'\lambda)\,\dot{x}'\ \!\!^{a}
\end{eqnarray}
for $u>0$. All geodesics starting from ${x'}^{a}$ will focus to the same point in the $x^{1}$-$x^{2}$ plane no matter what the initial $\,\dot{x}'\ \!\!^{a}$ is if 
\begin{eqnarray}
u-u'+uu'\lambda=0\Rightarrow\frac{1}{u}-\frac{1}{u'}=\lambda
\end{eqnarray}
This is just the lens equation in geometric optics with focal length $1/\lambda$. It is consistent with our previous result where geodesics with perpendicular incidence will all focus to a point at $u=1/\lambda$. For $u'<-1/\lambda$, this equation gives
\begin{eqnarray}
u_{c}=\frac{|u'|}{\lambda|u'|-1}>0
\end{eqnarray}
which is the location of the conjugate plane. On this conjugate plane, the geodesics focus to a point with coordinates $x_{c}^{a}=(1-u_{c}\lambda)\,x'\ \!\!^{a}$ and 
\begin{eqnarray}
v_{c}=v'+\frac{1}{2}\xi\lambda(1+u'\lambda)^{-1}u'\,^{2}-\frac{1}{2}\lambda(1+u'\lambda)^{-1}\sum_{a=1,2}x'\ \!\!^{a}x'\ \!\!^{a}.
\end{eqnarray}

In the nondegenerate case with $\lambda_{1}=-\lambda=-\lambda_{2}$, the conjugate plane is again located at $u_{c}=|u'|/(\lambda|u'|-1)$. From Eqs.~(\ref{xsolution1}) and (\ref{vsolution1}), we can obtain the image on this conjugate plane of the geodesics originating from ${x'}^{a}$ and with velocity $\dot{x}'$$^{a}$. Since the image should not depend on the choice of ${x'}^{a}$, we can simplify the expressions by taking ${x'}^{a}=0$. Then, $x_{c}^{1}=0$, $x_{c}^{2}=-2\lambda u'^{2}(1+\lambda u')^{-1}\dot{x}'$$^{2}$, and 
\begin{eqnarray}
v_{c}&=&v'+\frac{1}{2}\xi\lambda u'^{2}(1+\lambda u')^{-1}-\lambda u'^{2}(1-\lambda u')(1+\lambda u')^{-1}\dot{x}'\,^{2}\dot{x}'\,^{2}\nonumber\\
&=&v'+\frac{1}{2}\xi\lambda u'^{2}(1+\lambda u')^{-1}-\frac{1}{4}\left(\frac{1-\lambda^{2}u'^{2}}{\lambda u'^{2}}\right)(x_{c}^{2})^{2},
\end{eqnarray}
which is again a parabola in the $v$-$x^{2}$ plane as in the perpendicular incidence case.

\subsection{Bitensors: World function and van Vleck determinant}
Closely related to geodesics are various bitensors which are functions of two spacetime points. They are crucial to the construction of Green functions we shall discuss in some detail in the next section. The first bitensor we need to consider is the world function $\sigma(z,z')$  which is basically one-half of the squared geodesic distance between of the spacetime points $z^{\mu}=(u,v,x^{a})$ and $z'^{\mu}=(u',v',x'^{a})$. Another one also of interest to us is the van Vleck determinant $\Delta(z,z')$ which is the determinant of the second derivative of the world function \cite{HarDri12}.

To derive the bitensors, it is more convenient to express the geodesics in terms of the end points $z^{\mu}$ and $z'^{\mu}$. From our previous discussion, we note that the geodesics in the regions $u>0$ and $u<0$, that is, away from the shockwave, are straight lines. Hence, it is easy to see that for $0>u>u''>u'$, we have the geodesics
\begin{eqnarray}
x''^{a}&=&\left(\frac{u''-u'}{u-u'}\right)x^{a}+\left(\frac{u-u''}{u-u'}\right)x'^{a}\nonumber\\
v''&=&\left(\frac{u''-u'}{u-u'}\right)v+\left(\frac{u-u''}{u-u'}\right)v'\label{endptsol}
\end{eqnarray}
For $u>u''>u'>0$, the geodesics also have the same equations as above. 

For $u>0$ and $u'<0$, that is, across the shockwave, we must consider the geodesics before and after encountering the shock separately, taking in account the discontinuities in $\dot{x}'$$^{a}$ and $v$ as given in Eqs.~(\ref{velmemory}) and (\ref{vdiscon}), respectively. For $0>u''>u'$, using the solution in Eq.~(\ref{endptsol}),
\begin{eqnarray}
x''^{a}&=&-\frac{u''}{u'}(x_{0}^{a}-x'^{a})+x_{0}^{a}\nonumber\\
v''&=&-\frac{u''}{u'}(v_{0}-v')+v_{0}\label{endptgeo1}
\end{eqnarray}
where $v_{0}=v(0^{-})$, and for $u>u''>0$,
\begin{eqnarray}
x''^{a}&=&\frac{u''}{u}(x^{a}-x_{0}^{a})+x_{0}^{a}\nonumber\\
v''&=&\frac{u''}{u}\left[v-v_{0}-\frac{1}{2}\sum_{a=1,2}\lambda_{a}(x_{0}^{a})^{2}\right]+\left[v_{0}+\frac{1}{2}\sum_{a=1,2}\lambda_{a}(x_{0}^{a})^{2}\right]\label{endptgeo2}
\end{eqnarray}
$x_{0}^{a}$ can be expressed in terms of the end point coordinates by requiring that the derivative across the shock in Eqs.~(\ref{endptgeo1}) and (\ref{endptgeo2}) is consistent with that in Eq.~(\ref{velmemory}). The result is that
\begin{eqnarray}
x_{0}^{a}=(u-u'-uu'\lambda_{a})^{-1}(ux'^{a}-u'x^{a})\label{x0value}
\end{eqnarray}
Similarly, by noting that there is also a discontinuity $\dot{v}$ from Eq.~(\ref{vsolution}), $v_{0}$ can be expressed as
\begin{eqnarray}
v_{0}=\frac{uv'-u'v}{u-u'}+\sum_{a=1,2}\frac{\lambda_{a}x_{0}^{a}}{2(u-u')}\left[(-2u+u'+uu'\lambda_{a})x_{0}^{a}+2ux'^{a}\right]\label{v0value}
\end{eqnarray}

Now, we are ready to work out the first bitensor, the world function $\sigma(z,z')$, defined by
\begin{eqnarray}
\sigma(z,z')=\frac{1}{2}(u-u')\int_{u'}^{u}du''\,g_{\mu\nu}(z'')\dot{z}''^{\mu}\dot{z}''^{\nu}\label{defsigma}
\end{eqnarray}
For both $0>u>u'$ and $u>u'>0$, the spacetime is Minkowski, and 
\begin{eqnarray}
\sigma(z,z')=-(u-u')(v-v')+\frac{1}{2}\sum_{a=1,2}(x^{a}-x'^{a})(x^{a}-x'^{a})\label{Minsigma}
\end{eqnarray}
For $u>0$ and $u'<0$, we use Eq.~(\ref{endptgeo1}) to describe the goedesic in $0>u''>u'$ with
\begin{eqnarray}
g_{\mu\nu}\dot{z}''^{\mu}\dot{z}''^{\nu}=\frac{2}{u'}(v_{0}-v')+\frac{1}{u'^{2}}\sum_{a=1,2}(x_{0}^{a}-x'^{a})(x_{0}^{a}-x'^{a})
\end{eqnarray}
and in $u>u''>0$ with
\begin{eqnarray}
g_{\mu\nu}\dot{z}''^{\mu}\dot{z}''^{\nu}=-\frac{2}{u}\left[v-v_{0}-\frac{1}{2}\sum_{a=1,2}\lambda_{a}(x_{0}^{a})^{2}\right]+\frac{1}{u^{2}}\sum_{a=1,2}(x_{0}^{a}-x^{a})(x_{0}^{a}-x^{a})
\end{eqnarray}
Putting these results into Eq.~(\ref{defsigma}), we obtain the world function across the shock
\begin{eqnarray}
&&\sigma(z,z')\nonumber\\
&=&\frac{1}{2}(u-u')\left[\int_{u'}^{0}du''\,g_{\mu\nu}(z'')\dot{z}''^{\mu}\dot{z}''^{\nu}+\int_{0}^{u}du''\,g_{\mu\nu}(z'')\dot{z}''^{\mu}\dot{z}''^{\nu}\right]\nonumber\\
&=&\frac{1}{2}(u-u')\bigg\{-2(v-v')\nonumber\\
&&\hskip 20pt+\sum_{a=1,2}(u-u'-uu'\lambda_{a})^{-1}\left[(1-u'\lambda_{a})(x^{a})^{2}+(1+u\lambda_{a})(x'^{a})^{2}-2x^{a}x'^{a}\right]\bigg\}\label{finalsigma}
\end{eqnarray}
where we have substituted $x_{0}^{a}$ and $v_{0}$ by the end point coordinates as given in Eqs.~(\ref{x0value}) and (\ref{v0value}).

In terms of the world function, one can define another bitensor, the van Vleck determinant
\begin{eqnarray}
\Delta(z,z')=-\frac{{\rm det}[-\nabla_{\mu}\nabla_{\nu'}\sigma(z,z')]}{\sqrt{-{\rm det}g_{\alpha\beta}(z)}\sqrt{{\rm det}g_{\alpha'\beta'}(z')}}
\end{eqnarray}
In Minkowski spacetime, that is, for both $0>u>u'$ and $u>u'>0$, $\Delta(z,z')=1$. It is more interesting to look at the case across the shock with the world function given by Eq.~(\ref{finalsigma}). Then,
\begin{eqnarray}
&&\partial_{v}\partial_{u'}\sigma(z,z')=\partial_{u}\partial_{v'}\sigma(z,z')=1\nonumber\\
&&\partial_{v}\partial_{v'}\sigma(z,z')=\partial_{x^{1}}\partial_{x'^{2}}\sigma(z,z')=\partial_{x^{2}}\partial_{x'^{1}}\sigma(z,z')=0
=\partial_{v}\partial_{x'^{a}}\sigma(z,z')=\partial_{x^{a}}\partial_{v'}\sigma(z,z')=0\nonumber\\
&&\partial_{x^{a}}\partial_{x'^{a}}\sigma(z,z')=-(u-u')(u-u'-uu'\lambda_{a})^{-1}
\end{eqnarray}
Since for $u\neq 0$, det$g_{\alpha\beta}(z)=-1$, the van Vleck determinant for $u>0$ and $u'<0$ is just
\begin{eqnarray}
\Delta(z,z')&=&[\partial_{x^{1}}\partial_{x'^{1}}\sigma(z,z')][\partial_{x^{2}}\partial_{x'^{2}}\sigma(z,z')]\nonumber\\
&=&(u-u')^{2}(u-u'-uu'\lambda_{1})^{-1}(u-u'-uu'\lambda_{2})^{-1}\label{finalDelta}
\end{eqnarray}

For both the world function in Eq.~(\ref{finalsigma}) and the van Vleck determinant in Eq.~(\ref{finalDelta}), there are terms proportional to $(u-u'-uu'\lambda_{a})^{-1}$. Hence, these two bitensors will divergent when $u$ approaches the conjugate plane of $u'$. For example, in the degenerate case, $\lambda_{1}=\lambda_{2}=-\lambda$, near the conjugate plane with $u_{c}=|u'|/(\lambda|u'|-1)$,
\begin{eqnarray}
\sigma(z,z')&=&(u-u_{c})^{-1}\left[-\frac{\lambda u_{c}^{2}}{2(1-\lambda u_{c})}\right]\sum_{a=1,2}\left[(x^{a}-(1-\lambda u_{c})x'^{a}\right]^{2}\nonumber\\
&&+\left[\frac{\lambda u_{c}^{2}}{1-\lambda u_{c}}(v-v')+\frac{1}{2}\sum_{a=1,2}(x^{a}-(1-\lambda u_{c})x'^{a})^{2}+\frac{1}{2}\lambda^{2}u_{c}^{2}\sum_{a=1,2}(x'^{a})^{2}\right]\nonumber\\
&&-(u-u_{c})\left[(v-v')+\frac{1}{2}\lambda(1-\lambda u_{c})\sum_{a=1,2}(x'^{a})^{2}\right]\label{sigmalimit1}
\end{eqnarray}
and
\begin{eqnarray}
\Delta(z,z')=(u-u_{c})^{-2}(\lambda^{2}u_{c}^{4})+(u-u_{c})^{-1}(-2\lambda u_{c}^{2})(1-\lambda u_{c})+(1-\lambda u_{c})^{2}\label{Deltalimit1}
\end{eqnarray}
We see that in this case the world function diverges as $(u-u_{c})^{-1}$ and the van Vleck determinant as $(u-u_{c})^{-2}$ when $u$ approaches $u_{c}$.

In the nondegenerate case, with $\lambda_{1}=-\lambda=-\lambda_{2}$, similar expansions about $u=u_{c}$ give
\begin{eqnarray}
\sigma(z,z')&=&(u-u_{c})^{-1}\left[-\frac{\lambda u_{c}^{2}}{2(1-\lambda u_{c})}\left(x^{1}-(1-\lambda u_{c})x'^{1}\right)^{2}\right]\nonumber\\
&&+\bigg[\frac{\lambda u_{c}^{2}}{1-\lambda u_{c}}(v-v')+\frac{1}{2}\big(x^{1}-(1-\lambda u_{c})x'^{1}\big)^{2}+\frac{1}{2}\lambda^{2}u_{c}^{2}(x'^{1})^{2}\nonumber\\
&&\hskip 40pt +\frac{1-2\lambda u_{c}}{4(1-\lambda u_{c})}(x^{2})^{2}+\frac{1}{4}(1+\lambda u_{c})(x'^{2})^{2}-\frac{1}{2}x^{2}x'^{2}
\bigg]
\nonumber\\
&&+\cdots\label{sigmalimit2}
\end{eqnarray}
and
\begin{eqnarray}
\Delta(z,z')=(u-u_{c})^{-1}\left(-\frac{\lambda u_{c}^{2}}{2}\right)+\frac{1}{4}(3-2\lambda u_{c})+(u-u_{c})\left(-\frac{1}{8\lambda u_{c}^{2}}\right)+\cdots\label{Deltalimit2}
\end{eqnarray}
In this case, both the world function and the van Vleck determinant diverge as $(u-u_{c})^{-1}$ as $u$ approaches $u_{c}$.

\section{Wightman function: Within and beyond the normal neighborhood}
In the previous section we have considered the classical properties of impulsive plane wave spacetimes including the geodesics, conjugate planes and various bitensors. Here in this section we turn to examine the properties of a quantum scalar field in these spacetimes. We shall first calculate the mode functions corresponding to the in and the out vacua. Then we examine the Bogoliubov coefficients connecting these modes. In addition, using these mode functions we shall construct the Wightman function starting from which various two-point correlation functions can be derived.

\subsection{In- and out-mode functions}
Here we look at a minimally coupled massive scalar field $\phi$ in the impulsive plane wave spacetime \cite{Klimcik88}. From the metric in Eq.~(\ref{planemetric}), Klein-Gordon equation in this spacetime can be expressed as
\begin{eqnarray}
&&(\Box-m^{2})\phi(z)=0\nonumber\\
&\Rightarrow&\left[-2\frac{\partial}{\partial u}\frac{\partial}{\partial v}+\sum_{a=1,2}\frac{\partial}{\partial x^{a}}\frac{\partial}{\partial x^{a}}-f(\vec{x})\delta(u)\frac{\partial^{2}}{\partial v^{2}}-m^{2}\right]\phi(z)=0
\end{eqnarray}
Since the metric is independent of $v$, it is apparent that one can write the Fourier mode of $\phi$ as
$e^{-ik_{-}v}\psi_{k_{-}}(u,\vec{x})$. Then the mode equation becomes
\begin{eqnarray}
\left[2ik_{-}\frac{\partial}{\partial u}+\sum_{a=1,2}\frac{\partial}{\partial x^{a}}\frac{\partial}{\partial x^{a}}+f(\vec{x})\delta(u)k_{-}^{2}-m^{2}\right]\psi_{k_{-}}(u,\vec{x})=0\label{KKeqn}
\end{eqnarray}
In Minkowski spacetime, $f(\vec{x})=0$, and the Fourier mode is
\begin{eqnarray}
\psi_{k_{-}\vec{k}}(u,\vec{x})=N_{k_{-}}\ e^{-\frac{i}{2k_{-}}(\vec{k}^{2}+m^{2})u}e^{i\vec{k}\cdot\vec{x}}.\label{Minmode}
\end{eqnarray}
The normalization constant $N_{k_{-}}$ can be determined by the scalar product
\begin{eqnarray}
\langle\phi_{1},\phi_{2}\rangle=-i\int_{u}dv\prod_{a=1,2}dx^{a}(\phi_{1}\partial_{v}\phi^{*}_{2}-\phi^{*}_{2}\partial_{v}\phi_{1}).\label{inner}
\end{eqnarray}
Requiring that
$\langle\phi_{k_{-}\vec{k}},\phi_{k'_{-}\vec{k}'}\rangle=\delta(k_{-}-k'_{-})\delta(\vec{k}-\vec{k}')$
gives $N_{k_{-}\vec{k}}=[(2\pi)^{3}(2k_{-})]^{-1/2}$.

For impulsive plane wave spacetime, Eq.~(\ref{KKeqn}) can be rewritten as
\begin{eqnarray}
2ik_{-}\frac{\partial}{\partial u}({\rm ln}\psi)+\frac{1}{\psi}\sum_{a=1,2}\frac{\partial}{\partial x^{a}}\frac{\partial}{\partial x^{a}}\psi+f(\vec{x})\delta(u)k_{-}^{2}-m^{2}=0.
\end{eqnarray}
The presence of the delta function indicates that ln$\psi$ across the shockwave at $u=0$ is discontinuous. This discontinuity can be obtained by integrating over $u\sim(-\epsilon,\epsilon)$ and then taking the limit $\epsilon\rightarrow 0$. 
\begin{eqnarray}
&&{\rm ln}\psi_{k_{-}}(0^{+}\vec{x})-{\rm ln}\psi_{k_{-}}(0^{-},\vec{x})=\frac{i}{2}k_{-}f(\vec{x})\nonumber\\
&\Rightarrow&\psi_{k_{-}}(0^{+},\vec{x})=e^{\frac{i}{2}k_{-}f(\vec{x})}\psi_{k_{-}}(0^{-},\vec{x})\label{psidiscon}
\end{eqnarray}
For an in-mode in this impulsive spacetime, the mode function is that of the Minkowski mode for $u<0$ before the interaction with the shockwave. That is, $\psi^{\rm in}_{k_{-}\vec{k}}(0^{-},\vec{x})=N_{k_{-}}e^{i\vec{k}\cdot\vec{x}}$. Hence, from Eq.~(\ref{psidiscon}), we have $\psi^{\rm in}_{k_{-}\vec{k}}(0^{+},\vec{x})=N_{k_{-}}e^{i\vec{k}\cdot\vec{x}}e^{\frac{i}{2}k_{-}f(\vec{x})}$. This can be used as an initial condition to work out $\psi^{\rm in}_{k_{-}\vec{k}}(u,\vec{x})$ for $u>0$. Take the Fourier transform
\begin{eqnarray}
\psi^{\rm in}_{k_{-}\vec{k}}(u,\vec{x})=\int\frac{d^{2}k'}{2\pi}e^{i\vec{k}'\cdot\vec{x}}\tilde{\psi}^{\rm in}_{k_{-}\vec{k}}(u,\vec{k}'),
\end{eqnarray}
for $u>0$. Putting this into Eq.~(\ref{KKeqn}), we have
\begin{eqnarray}
\left[2ik_{-}\frac{\partial}{\partial u}-\vec{k}'^{2}-m^{2}\right]\tilde{\psi}^{\rm in}_{k_{-}\vec{k}}(u,\vec{k}')=0,
\end{eqnarray}
with the initial condition
\begin{eqnarray}
\tilde{\psi}^{\rm in}_{k_{-}\vec{k}}(0,\vec{k}')=N_{k_{-}}\int\frac{d^{2}x'}{2\pi}e^{i(\vec{k}-\vec{k}')\cdot\vec{x}'}e^{\frac{i}{2}k_{-}f(\vec{x}')}
\end{eqnarray}
The solution to this is just, for $u>0$,
\begin{eqnarray}
\tilde{\psi}^{\rm in}_{k_{-}\vec{k}}(u,\vec{k}')=N_{k_{-}}e^{-\frac{i}{2k_{-}}(\vec{k}'^{2}+m^{2})u}\int\frac{d^{2}x'}{2\pi}e^{i(\vec{k}-\vec{k}')\cdot\vec{x}'}e^{\frac{i}{2}k_{-}f(\vec{x}')}\label{inmode}
\end{eqnarray}
Finally, we have the in-mode function $\phi^{\rm in}_{k_{-}\vec{k}}(u,v,\vec{x})$ \cite{GKMTT21}. For $u<0$, we have from Eq.~(\ref{Minmode}) the in-mode
\begin{eqnarray}
\phi^{\rm in}_{k_{-}\vec{k}}(z)=N_{k_{-}}e^{-ik_{-}v}e^{-\frac{i}{2k_{-}}(\vec{k}^{2}+m^{2})u}e^{i\vec{k}\cdot\vec{x}}.\label{inmodesmaller}
\end{eqnarray}
which is the same as the Minkowski mode function. For $u>0$, we have from Eq.~(\ref{inmode}),
\begin{eqnarray}
\phi^{\rm in}_{k_{-}\vec{k}}(z)=N_{k_{-}}e^{-ik_{-}v}e^{i\vec{k}\cdot\vec{x}}\int\frac{d^{2}x'\,d^{2}k'}{(2\pi)^{2}}e^{-i(\vec{k}-\vec{k}')\cdot(\vec{x}-\vec{x}')}e^{\frac{i}{2}k_{-}f(\vec{x}')}e^{-\frac{i}{2k_{-}}(\vec{k}'^{2}+m^{2})u}\label{inmodelarger}
\end{eqnarray}

In a similar fashion, we can also obtain the out-mode function $\phi^{\rm out}_{k_{-}\vec{k}}(z)$. This out-mode function is just the Minkowski mode function for $u>0$.  The delta function term in Eq.~(\ref{KKeqn}) is then used to derive the discontinuity of $\psi_{k_{-}\vec{k}}$ at $u=0$. This discontinuity gives a boundary condition to calculate the mode function for $u<0$. The result is that for $u<0$,
\begin{eqnarray}
\phi^{\rm out}_{k_{-}\vec{k}}(z)=N_{k_{-}}e^{-ik_{-}v}e^{i\vec{k}\cdot\vec{x}}\int\frac{d^{2}x'\,d^{2}k'}{(2\pi)^{2}}e^{-i(\vec{k}-\vec{k}')\cdot(\vec{x}-\vec{x}')}e^{-\frac{i}{2}k_{-}f(\vec{x}')}e^{-\frac{i}{2k_{-}}(\vec{k}'^{2}+m^{2})u},\label{outmodesmaller}
\end{eqnarray}
while for $u>0$,
\begin{eqnarray}
\phi^{\rm out}_{k_{-}\vec{k}}(z)=N_{k_{-}}e^{-ik_{-}v}e^{-\frac{i}{2k_{-}}(\vec{k}^{2}+m^{2})u}e^{i\vec{k}\cdot\vec{x}}.
\end{eqnarray}

\subsection{Focusing of modes}
In our previous consideration of geodesics, we have encountered the phenomenon of geodesic focusing at the conjugate plane. For example, geodesics with perpendicular incidence in spacetimes with $f(\vec{x})=\sum_{a=1,2}\lambda_{a}(x^{a})^{2}$ will focus to a point in the $x^{1}$-$x^{2}$ plane in the degenerate case with $\lambda_{1}=\lambda_{2}=-\lambda$ or to a line in the nondegenerate case with $\lambda_{1}=-\lambda=-\lambda_{2}$ at focal point $u=1/\lambda$. In fact, similar focusing effect also occur for the in- and out-modes that we have just examined \cite{GarVer91}. To explore this phenomenon, we first further simplify the mode functions for $f(\vec{x})=\sum_{a=1,2}\lambda_{a}(x^{a})^{2}$.

With this form of $f(\vec{x})$, the integrals over $\vec{k}'$ and $\vec{x}'$ in Eq.~(\ref{inmodelarger}) are all gaussian. To render the integrals convergent, we put in an infinitesimal negative imaginary part to $u$, that is, $u\rightarrow u-i\epsilon$. Then, one can use the formula
\begin{eqnarray}
\int dx\,e^{iax^{2}+ibx}=\sqrt{\frac{i\pi}{a}}e^{-ib^{2}/4a},
\end{eqnarray}
where $a$ and $b$ are constants with Im $a>0$, to arrive at the in-mode for $u>0$,
\begin{eqnarray}
\phi^{\rm in}_{k_{-}\vec{k}}(z)&=&N_{k_{-}}e^{-k_{-}v}e^{-\frac{im^{2}}{2k_{-}}(u-i\epsilon)}e^{\frac{ik_{-}}{2(u-i\epsilon)}\vec{x}^{2}}\nonumber\\
&&\ \ \prod_{a=1,2}\left[1+\lambda_{a}(u-i\epsilon)\right]^{-1/2}e^{-\frac{i}{2k_{-}}\left(\frac{1}{u-i\epsilon}+\lambda_{a}\right)^{-1}\left(k^{a}-\frac{k_{-}x^{a}}{u-i\epsilon}\right)^{2}}
\end{eqnarray}
Similarly, we can also work out the out-mode function for $u<0$.
\begin{eqnarray}
\phi^{\rm out}_{k_{-}\vec{k}}(z)&=&N_{k_{-}}e^{-k_{-}v}e^{-\frac{im^{2}}{2k_{-}}(u-i\epsilon)}e^{\frac{ik_{-}}{2(u-i\epsilon)}\vec{x}^{2}}\nonumber\\
&&\ \ \prod_{a=1,2}\left[1-\lambda_{a}(u-i\epsilon)\right]^{-1/2}e^{-\frac{i}{2k_{-}}\left(\frac{1}{u-i\epsilon}-\lambda_{a}\right)^{-1}\left(k^{a}-\frac{k_{-}x^{a}}{u-i\epsilon}\right)^{2}}
\end{eqnarray}

Now, we can examine the focusing effect of these mode functions. In the degenerate case with $\lambda_{1}=\lambda_{2}=-\lambda$, the in-mode for $u>0$ becomes
\begin{eqnarray}
\phi^{\rm in}_{k_{-}\vec{k}}(z)
=N_{k_{-}}e^{-k_{-}v}e^{-\frac{im^{2}}{2k_{-}}(u-i\epsilon)}e^{\frac{ik_{-}}{2(u-i\epsilon)}\vec{x}^{2}}
\left[1-\lambda(u-i\epsilon)\right]^{-1}\prod_{a=1,2}}e^{-\frac{i}{2k_{-}}\left(\frac{1}{u-i\epsilon}-\lambda\right)^{-1}\left(k^{a}-\frac{k_{-}x^{a}}{u-i\epsilon}\right)^{2}\nonumber\\
\end{eqnarray}
We can immediately see that at the focal point $u=1/\lambda$, $\phi^{\rm in}_{k_{-}\vec{k}}$ is singular as $\epsilon\rightarrow 0$ due to the presence of terms like $(1-\lambda(u-i\epsilon))^{-1}$. Let us examine this expression more closely. Take $u=1/\lambda$, and we have
\begin{eqnarray}
\left.\phi^{\rm in}_{k_{-}\vec{k}}(z)\right|_{u=1/\lambda}
=\lim_{\epsilon\rightarrow 0}N_{k_{-}}e^{-k_{-}v}e^{-\frac{im^{2}}{2k_{-}\lambda}}e^{\frac{i}{2}k_{-}\lambda\vec{x}^{2}}
(i\lambda\epsilon)^{-1}\prod_{a=1,2}}e^{-\frac{1}{2k_{-}\lambda^{2}\epsilon}\left(k^{a}-k_{-}\lambda x^{a}\right)^{2}\label{inmodesing}
\end{eqnarray}
The singularity in this expression as $\epsilon\rightarrow 0$ can be described by the delta function since one has
\begin{eqnarray}
\lim_{\epsilon\rightarrow 0}\frac{1}{\sqrt{\epsilon}}e^{-\frac{1}{2k_{-}\lambda^{2}\epsilon}\left(k^{a}-k_{-}\lambda x^{a}\right)^{2}}
&=&\sqrt{\frac{2\pi}{k_{-}}}\,\delta\left(x^{a}-\frac{k^{a}}{k_{-}\lambda}\right)\label{deltalimit}
\end{eqnarray}
Therefore, 
\begin{eqnarray}
\left.\phi^{\rm in}_{k_{-}\vec{k}}\right|_{u=1/\lambda}
=N_{k_{-}}\left(\frac{-i2\pi}{k_{-}\lambda}\right)e^{-ik_{-}v}e^{\frac{i}{2k_{-}\lambda}(\vec{k}^{2}-m^{2})}\prod_{a=1,2}\delta\left(x^{a}-\frac{k^{a}}{k_{-}\lambda}\right)\label{inmodedelta}
\end{eqnarray}
At the plane with $u=1/\lambda$ the focal length, the in-mode function focuses in such a way that it is non-zero only when $x^{a}=k^{a}/k_{-}\lambda$, $a=1,2$, which is a point on the $x^{1}$-$x^{2}$ plane. On this point the mode function is described by a two-dimensional delta function. Actually, a direct evaluation of the integrals in Eq.~(\ref{inmodelarger}) while putting $u=1/\lambda$ will also produce Eq.~(\ref{inmodedelta}) with the delta functions. 

For the out-mode, focusing occurs at $u=-1/\lambda$, giving
\begin{eqnarray}
\left.\phi^{\rm out}_{k_{-}\vec{k}}\right|_{u=-1/\lambda}
=N_{k_{-}}\left(\frac{i2\pi}{k_{-}\lambda}\right)e^{-k_{-}v}e^{-\frac{i}{2k_{-}\lambda}(\vec{k}^{2}-m^{2})}\prod_{a=1,2}\delta\left(x^{a}+\frac{k^{a}}{k_{-}\lambda}\right)
\end{eqnarray}

Next, we consider the nondegenerate case with $\lambda_{1}=-\lambda=-\lambda_{2}$ in which focusing of modes also happens. This time the mode function focuses to a line in the $x^{1}$-$x^{2}$ plane instead of a point in the previous case. In particular, at $u=1/\lambda$, the in-mode
\begin{eqnarray}
&&\left.\phi^{\rm in}_{k_{-}\vec{k}}(z)\right|_{u=1/\lambda}\nonumber\\
&=&N_{k_{-}}\sqrt{\frac{-i\pi}{k_{-}\lambda}}e^{-ik_{-}v}e^{\frac{i}{2k_{-}\lambda}((k^{1})^{2}-m^{2})}e^{-\frac{i}{4k_{-}\lambda}[(k^{2}-k_{-}\lambda x^{2})^{2}-2k_{-}^{2}\lambda^{2}(x^{2})^{2}]}\delta\left(x^{1}-\frac{k^{1}}{k_{-}\lambda}\right)
\end{eqnarray}
where we have an one-dimensional delta function this time. For the out-mode at $u=-1/\lambda$,
\begin{eqnarray}
&&\left.\phi^{\rm out}_{k_{-}\vec{k}}(z)\right|_{u=-1/\lambda}\nonumber\\
&=&N_{k_{-}}\sqrt{\frac{i\pi}{k_{-}\lambda}}e^{-ik_{-}v}e^{-\frac{i}{2k_{-}\lambda}((k^{1})^{2}-m^{2})}e^{\frac{i}{4k_{-}\lambda}[(k^{2}+k_{-}\lambda x^{2})^{2}-2k_{-}^{2}\lambda^{2}(x^{2})^{2}]}\delta\left(x^{1}+\frac{k^{1}}{k_{-}\lambda}\right)\nonumber\\
\end{eqnarray}

\subsection{Bogoliubov coefficients}
In the impulsive plane wave spacetimes, we have developed two sets of mode functions $\phi^{\rm in}_{k_{-}\vec{k}}(z)$ and $\phi^{\rm out}_{k_{-}\vec{k}}(z)$. Here, we examine the Bogoliubov transformations between them \cite{GarVer91}. 
\begin{eqnarray}
\phi^{\rm out}_{l_{-}\vec{l}}(z)=\int_{0}^{\infty}dk_{-}\int\,d^{2}k\left[\alpha_{l_{-}\vec{l},k_{-}\vec{k}}\,\phi^{\rm in}_{k_{-}\vec{k}}(z)+\beta_{l_{-}\vec{l},k_{-}\vec{k}}\,\phi^{\rm in\, *}_{k_{-}\vec{k}}(z)
\right]
\end{eqnarray}
The Bogoliubov coefficients $\alpha_{l_{-}\vec{l},k_{-}\vec{k}}$ and $\beta_{l_{-}\vec{l},k_{-}\vec{k}}$ are given by
\begin{eqnarray}
\alpha_{l_{-}\vec{l},k_{-}\vec{k}}&=&\langle\phi^{\rm out}_{l_{-}\vec{l}}(z),\phi^{\rm in}_{k_{-}\vec{k}}(z)\rangle\ \ \ ;\ \ \ \beta_{l_{-}\vec{l},k_{-}\vec{k}}=-\langle\phi^{\rm out}_{l_{-}\vec{l}}(z),\phi^{\rm in\,*}_{k_{-}\vec{k}}(z)\rangle
\end{eqnarray}
where the inner product is as defined in Eq.~(\ref{inner}).

We first look at the coefficient $\beta_{l_{-}\vec{l},k_{-}\vec{k}}$ because 
$\sum_{k_{-}\vec{k}}|\beta_{l_{-}\vec{l},k_{-}\vec{k}}|^{2}$ is related to the spectrum of the particle produced starting with the in-vacuum. By definition the inner product is independent of the constant $u$ plane chosen to evaluate it. If we take $u<0$, then $\phi^{\rm in}_{k_{-}\vec{k}}(z)$ is given by Eq.~(\ref{inmodesmaller}) and $\phi^{\rm out}_{l_{-}\vec{l}}(z)$ by Eq.~(\ref{outmodesmaller}). 
\begin{eqnarray}
\beta_{l_{-}\vec{l},k_{-}\vec{k}}&=&\int dv\int d^{2}x\, N_{l_{-}}N_{k_{-}}\ (k_{-}-l_{-})\,e^{-i(l_{-}+k_{-})v}e^{i(\vec{l}+\vec{k})\cdot\vec{x}}e^{-\frac{i}{2k_{-}}(\vec{k}^{2}+m^{2})u}\nonumber\\
&&\ \ \int\frac{d^{2}x'\,d^{2}l'}{(2\pi)^{2}}e^{-i(\vec{l}-\vec{l}')\cdot(\vec{x}-\vec{x}')}e^{-\frac{i}{2l_{-}}(\vec{l}'^{2}+m^{2})u}e^{-\frac{i}{2}l_{-}f(\vec{x}')}
\end{eqnarray}
Integrating over $v$, $\vec{x}$, and $\vec{l}'$, we have
\begin{eqnarray}
\beta_{l_{-}\vec{l},k_{-}\vec{k}}=N_{l_{-}}N_{k_{-}}2\pi (k_{-}-l_{-})\,\delta(k_{-}+l_{-})\int\,d^{2}x'\,e^{i(\vec{k}+\vec{l})\cdot\vec{x}'+\frac{i}{2}k_{-}f(\vec{x}')}
\end{eqnarray}
which is indeed independent of $u$. 
Since both momenta $k_{-},l_{-}\geq 0$, we have $k_{-}+l_{-}\geq 0$. The presence of the delta function $\delta(k_{-}+l_{-})$ would require $\beta_{l_{-}\vec{l},k_{-}\vec{k}}=0$. This means that  there is no particle production in the impulsive plane wave spacetime. The positive frequency in-modes will not mix with the negative frequency out-modes and one can identify the in- and the out-vacua.

Although $\beta_{l_{-}\vec{l},k_{-}\vec{k}}$ vanishes, $\alpha_{l_{-}\vec{l},k_{-}\vec{k}}$ does not. To see that we again take $u<0$, and
\begin{eqnarray}
\alpha_{l_{-}\vec{l},k_{-}\vec{k}}=|N_{k_{-}}|^{2}(4\pi k_{-})\delta(k_{-}-l_{-})\int\,d^{2}x'\,e^{-i(\vec{k}-\vec{l})\cdot\vec{x}'-\frac{i}{2}k_{-}f(\vec{x}')}
\end{eqnarray}
For $f(\vec{x})=\sum_{a=1,2}\lambda_{a}(x^{a})^{2}$, we can further simplify the expression giving
\begin{eqnarray}
\alpha_{l_{-}\vec{l},k_{-}\vec{k}}=\left(\frac{1}{2\pi i k_{-}}\right)\delta(k_{-}-l_{-})\prod_{a=1,2}(\lambda_{a})^{-1/2}e^{\frac{i}{2k_{-}\lambda_{a}}(k^{a}-l^{a})^{2}}
\end{eqnarray}
The nonzero value of this Bogoliubov coefficient indicates that although the positive frequency in-modes of do not mix with the negative frequency out-modes, they do mix with the positive frequency ones. This comes about due to the interaction of the quantum field with the shockwave at $u=0$.

\subsection{Wightman functions in impulsive plane wave spacetime}\label{wightman}
After the development of complete sets of mode functions, we are in the position to use them to construct various kinds of two-point functions. These two-point functions will be crucial in the calculation and understanding of different quantum processes. The most basic one would be the Wightman function which is defined by the expectation value of two quantum field operators at different spacetime points \cite{DeWBre60,BirDav82}. The positive Wightman is given by $G_{+}(z,z')=\langle\phi(z)\phi(z')\rangle$, while the negative Wightman function $G_{-}(z,z')=\langle\phi(z')\phi(z)\rangle$. For a real scalar field, $G_{-}(z,z')=G_{+}(z,z')^{*}$. Hence, both the Pauli-Jordan function $G(z,z')=-i\langle[\phi(z),\phi(z')]\rangle$ and the Hadamard elementary function $G^{(1)}(z,z')=\langle\{\phi(z),\phi(z')\}\rangle$ are real. This is also true for the retarded Green function $G_{R}(z,z')=-\theta(u-u')G(z,z')$ and the advanced Green function $G_{A}(z,z')=\theta(u'-u)G(z,z')$. The Wightman function will be important for our consideration of the noise kernel in the next section. The retarded Green function is, of course, crucial in discussing initial value problems.

In terms of the mode functions, the Wightman function can be expressed as
\begin{eqnarray}
G_{+}(z,z')=\int_{0}^{\infty}dk_{-}\int\,d^{2}k\,\phi^{\rm in}_{k_{-}\vec{k}}(z)\,\phi^{in\, *}_{k_{-}\vec{k}}(z').\label{intermsofmode}
\end{eqnarray}
For $u,u'<0$, we have the Minkowski Wightman function
\begin{eqnarray}
G_{+}(z,z')&=&\int_{0}^{\infty}\frac{dk_{-}}{2k_{-}}\int\,\frac{d^{2}k}{(2\pi)^{3}}e^{-ik_{-}(v-v')}e^{i\vec{k}\cdot(\vec{x}-\vec{x}')}e^{-\frac{i}{2k_{-}}(\vec{k}^{2}+m^{2})(u-u')}\nonumber\\
&=&\frac{-i}{8\pi^{2}(u-u')}\int_{0}^{\infty}dk_{-}e^{-ik_{-}(v-v')}e^{-\frac{im^{2}}{2k_{-}}(u-u')}e^{\frac{ik_{-}}{2(u-u')}(\vec{x}-\vec{x}')^{2}}\label{MinWightman}
\end{eqnarray}
We shall work out this Minkowski Wightman function in some detail \cite{BogShi80} because the Wightman function with other spacetime points can be analyzed in the same fashion. The integral over $k_{-}$ above is singular. To make this singularity more transparent, we first take a derivative with respect to $d\equiv(\vec{x}-\vec{x}')^{2}$. Hence, we write $G_{+}(z,z')=\partial F(d)/\partial d$, where
\begin{eqnarray}
F(d)=-\frac{1}{4\pi^{2}}\int_{0}^{\infty}\frac{dk_{-}}{k_{-}}e^{-ik_{-}(v-v')}e^{-\frac{im^{2}}{2k_{-}}(u-u')}e^{\frac{ik_{-}d}{2(u-u')}}
\end{eqnarray}
where $F(d)$ is a convergent integral. 

To continue we consider the $u>u'$ and $u<u'$ cases separately. For $u>u'$, we define $T=(u-u')/2k_{-}$, and 
$F(d)=-(1/4\pi^{2})\int_{0}^{\infty}dT\,{T}^{-1}e^{-im^{2}T+i\sigma/2T}$,
where $\sigma=-(u-u')(v-v')+d/2$ is just the world function we have defined previously. For $u<u'$, we need to define $T=-(u-u')/2k_{-}$ for $T$ to be positive. Then,
$F(d)=-(1/4\pi^{2})\int_{0}^{\infty}dT\,{T}^{-1}e^{im^{2}T-i\sigma/2T}$
which is the complex conjugate of the function for $u>u'$.

Next, we need to distinguish cases with $\sigma>0$ and $\sigma<0$, that is, cases with spacelike and timelike separations, respectively. We shall use the formulas \cite{GraRyz07}, for $a,b>0$,
\begin{eqnarray}
&&\int_{0}^{\infty}\frac{dx}{x}\sin\left(a^{2}x+\frac{b^{2}}{x}\right)=\pi J_{0}(2ab)\ \ \ ;\ \ \ 
\int_{0}^{\infty}\frac{dx}{x}\cos\left(a^{2}x+\frac{b^{2}}{x}\right)=-\pi N_{0}(2ab)\nonumber\\
&&\int_{0}^{\infty}\frac{dx}{x}\sin\left(a^{2}x-\frac{b^{2}}{x}\right)=0\ \ \ ;\ \ \ 
\int_{0}^{\infty}\frac{dx}{x}\cos\left(a^{2}x-\frac{b^{2}}{x}\right)=2 K_{0}(2ab)
\end{eqnarray}
where $J_{n}(x)$ and $N_{n}(x)$ are Bessel functions of the first and second kind, respectively, and $K_{n}(x)$ is the modified Bessel function. Then, we have for $u>u'$ and $\sigma<0$,
\begin{eqnarray}
F(d)&=&-\frac{1}{4\pi^{2}}\int_{0}^{\infty}\frac{dT}{T}\left[\cos\left(m^{2}T-\frac{\sigma}{2T}\right)-i\sin\left(m^{2}T-\frac{\sigma}{2T}\right)\right]\nonumber\\
&=&\frac{1}{4\pi}\left[iJ_{0}(m\sqrt{-2\sigma})+N_{0}(m\sqrt{-2\sigma})\right]\label{fdssmall}
\end{eqnarray}
For $u<u'$ and $\sigma<0$, we have $F(d)=[-iJ_{0}(m\sqrt{-2\sigma})+N_{0}(m\sqrt{-2\sigma})]/4\pi$, the complex conjugate of Eq.~(\ref{fdssmall}). For $\sigma>0$, we have $F(d)=-K_{0}(m\sqrt{2\sigma})/2\pi^{2}$ which is real so the expressions for $u>u'$ and $u<u'$ are the same.

Collectively, using the step functions, one can express $F(d)$ for all $u$, $u'$ and $\sigma$ as
\begin{eqnarray}
F(d)&=&\frac{1}{4\pi}\theta(u-u')\theta(-\sigma)\left[iJ_{0}(m\sqrt{-2\sigma})+N_{0}(m\sqrt{-2\sigma})\right]\nonumber\\
&&\ \ +\frac{1}{4\pi}\theta(u'-u)\theta(-\sigma)\left[-iJ_{0}(m\sqrt{-2\sigma})+N_{0}(m\sqrt{-2\sigma})\right]\nonumber\\
&&\ \ +\theta(\sigma)\left(-\frac{1}{2\pi^{2}}\right)K_{0}(m\sqrt{2\sigma}).
\end{eqnarray}
Now, the positive Wightman function is given by the derivative of $F(d)$.
We note that $\partial\theta(\pm\sigma)/\partial d=\pm\delta(\sigma)/2$, and we have finally for $u,u'<0$
\begin{eqnarray}
G_{+}(z,z')&=&-\frac{i}{8\pi}\epsilon(u-u')\delta(\sigma)+\frac{im}{8\pi\sqrt{-2\sigma}}\epsilon(u-u')\theta(-\sigma)J_{1}(m\sqrt{-2\sigma})\nonumber\\
&&\ \ +\frac{m}{8\pi\sqrt{-2\sigma}}\theta(-\sigma)N_{1}(m\sqrt{-2\sigma})+\frac{m}{4\pi^{2}\sqrt{2\sigma}}\theta(\sigma)K_{1}(m\sqrt{2\sigma})\label{finalMinWightman}
\end{eqnarray}
where $\epsilon(u-u')=\theta(u-u')-\theta(u'-u)$. 

For $u,u'>0$, the in-mode function is given by Eq.~(\ref{inmodelarger}). The positive Wightman is then given by Eq.~(\ref{intermsofmode}). After the integration over $\vec{k}$, the terms with $f(\vec{x})$ just cancelled each other, and the resulting expression is the same as that in Minkowski spacetime given by Eq.~(\ref{MinWightman}).

We now come to the more interesting case with $u>0$ and $u'<0$, that is, across the shock. Then, from the mode functions in Eqs~(\ref{inmodesmaller}) and (\ref{inmodelarger}),
\begin{eqnarray}
G_{+}(z,z')&=&\int_{0}^{\infty}\frac{dk_{-}}{2k_{-}}\,e^{-ik_{-}(v-v')}e^{-\frac{im^{2}}{2k_{-}}(u-u')}\int\,d^{2}x''\,e^{\frac{i}{2}k_{-}f(\vec{x}'')}\nonumber\\
&&\ \ \int\,\frac{d^{2}k}{(2\pi)^{2}}e^{\frac{iu}{2k_{-}}\vec{k}^{2}}e^{-i\vec{k}\cdot(\vec{x}'-\vec{x}'')}\int\frac{d^{2}k''}{(2\pi)^{2}}e^{-\frac{iu}{2k_{-}}\vec{k}''^{2}}e^{i\vec{k}''\cdot(\vec{x}-\vec{x}'')}
\end{eqnarray}
For $f(\vec{x})=\sum_{a=1,2}\lambda_{a}(x^{a})^{2}$, the integrations over $\vec{k}$, $\vec{k}''$ and $\vec{x}''$ are all gaussian and can be done readily. Hence, we have
\begin{eqnarray}
G_{+}(z,z')&=&-\frac{i}{8\pi^{2}}\int_{0}^{\infty}dk_{-}e^{-ik_{-}(v-v')}e^{-\frac{im^{2}}{2k_{-}}(u-u')}\nonumber\\
&&\ \ \prod_{a=1,2}(u-u'-\lambda_{1})^{-1/2}e^{\frac{ik_{-}}{2}(u-u'-uu'\lambda_{a})^{-1}\left[(1-u'\lambda_{a})(x^{a})^{2}-(1-u\lambda_{a})(x'^{a})^{2}-2x^{a}x'^{a})\right]}\nonumber\\
\end{eqnarray}
Because $u>u'$, we can set $T=(u-u')/2k_{-}$ which is positive. So the positive Wightman function becomes
\begin{eqnarray}
&&G_{+}(z,z')\nonumber\\
&=&-\frac{i}{16\pi^{2}}\left[(u-u')^{2}(u-u-uu'\lambda_{1})^{-1}(u-u'-uu'\lambda_{2})^{-1}\right]^{1/2}\nonumber\\
&&\ \ \int_{0}^{\infty}\frac{dT}{T^{2}}e^{-im^{2}T}e^{\frac{i}{2T}(u-u')\left\{-(v-v')+\frac{1}{2}\sum_{a=1,2}(u-u'-uu'\lambda_{a})^{-1}\left[(1-u'\lambda_{a})(x^{a})^{2}+(1+u\lambda_{a})(x'^{a})^{2}-2x^{a}x'^{a})\right]\right\}}\nonumber\\
\end{eqnarray}
From Eq.~(\ref{finalDelta}), we immediate recognize that the prefactor in front of the integral is just the square root of the van Vleck determinant $\Delta(z,z')$. Also, from Eq.~(\ref{finalsigma}), the expression in the curly bracket is related to the world function $\sigma(z,z')$. Therefore, the positive Wightman function can be written compactly as
\begin{eqnarray}
G_{+}(z,z')=-\frac{i}{16\pi^{2}}\sqrt{\Delta(u,u')}\int_{0}^{\infty}\frac{dT}{T^{2}}e^{-im^{2}T}e^{i\sigma(z,z')/2T}\label{Wightmanacross}
\end{eqnarray}
This expression is also valid for $0>u>u'$ and $u>u'>0$. In those cases, $\Delta(u,u')=1$ and $\sigma(z,z')$ has the Minkowski value as in Eq.~(\ref{Minsigma}). Then, Eq.~(\ref{Wightmanacross}) reduces to that in Minkowski spacetime. Using the result in Eq.~(\ref{finalMinWightman}), we immediately have the final expression for the positive Wightman function valid for all $u$ and $u'$.
\begin{eqnarray}
G_{+}(z,z')&=&\sqrt{\Delta(u,u')}\bigg[-\frac{i}{8\pi}\epsilon(u-u')\delta(\sigma)+\frac{im}{8\pi\sqrt{-2\sigma}}\epsilon(u-u')\theta(-\sigma)J_{1}(m\sqrt{-2\sigma})\nonumber\\
&&\hskip 50pt +\frac{m}{8\pi\sqrt{-2\sigma}}\theta(-\sigma)N_{1}(m\sqrt{-2\sigma})+\frac{m}{4\pi^{2}\sqrt{2\sigma}}\theta(\sigma)K_{1}(m\sqrt{2\sigma})\bigg]\label{finalWightman}
\end{eqnarray}

As we have found that both the world function and the van Vleck determinant diverge when they approach the conjugate plane, the above expression is valid only within these planes. In such a normal neighborhood, using the series expansions of the Bessel functions, one can develop a Hadamard form for the positive Wightman functions \cite{DeWBre60,DecFol08}.
\begin{eqnarray}
G_{+}(z,z')=\frac{1}{8\pi^{2}}\left[\frac{U(z,z')}{\sigma+ i(u-u')\epsilon}+V(z,z')\,{\rm ln}\left(\frac{m^{2}\sigma}{2}+ i(u-u')\epsilon\right)+W(z,z')\right]\label{Hadamard}
\end{eqnarray}
as $\epsilon\rightarrow 0$, where we have used the formulas 
\begin{eqnarray}
\frac{1}{\sigma+ i\epsilon}={\cal P}\frac{1}{\sigma}- i\pi\delta(\sigma)\ \ \ ;\ \ \ {\rm ln}(\sigma+ i\epsilon)={\rm ln}|\sigma|+ i\pi\theta(-\sigma)
\end{eqnarray}
where ${\cal P}$ means the principal value. The functions $V(z,z')$, $U(z,z')$ and $W(z,z')$ can be expressed as power series of $\sigma$ as
\begin{eqnarray}
U(z,z')&=&\sqrt{\Delta(u,u')}\nonumber\\
V(z,z')&=&\frac{m^{2}\sqrt{\Delta(u,u')}}{2}\sum_{n=0}^{\infty}\frac{1}{n!(n+1)!}\left(\frac{m\sqrt{2\sigma}}{2}\right)^{2n}\nonumber\\
W(z,z')&=&-\frac{m^{2}\sqrt{\Delta(u,u')}}{2}\sum_{n=0}^{\infty}\frac{1}{n!(n+1)!}\left(\psi(n+1)+\psi(n+2)\right)\left(\frac{m\sqrt{2\sigma}}{2}\right)^{2n}
\end{eqnarray}
From this, we also see that as $m\rightarrow 0$, that is, for the massless case,
\begin{eqnarray}
G_{+}(z,z')=\frac{\sqrt{\Delta(u,u')}}{8\pi^{2}}\left[\frac{1}{\sigma}-i\pi \epsilon(u-u')\delta(\sigma)\right].
\end{eqnarray}
where $\epsilon(u-u')=\theta(u-u')-\theta(u'-u)$ that we have defined before.

\subsection{Two-point functions beyond the normal neighborhood}
As we have mentioned in the last subsection, the expressions in Eqs.~(\ref{finalWightman}) and (\ref{Hadamard}) are no longer valid when $u$ approaches the conjugate plane where both $\sigma(z,z')$ and $\Delta(u,u')$ diverge. However, the mode functions we work out earlier are actually valid for all spacetime points. Therefore, we expect that they can also be used to construct the two-point functions, notably the Wightman function, with any spacetime points, even for those on the conjugate plane. These two-point functions are called global two-point functions because they are defined everywhere even beyond the normal neighborhood. Recently, there is much interest in understanding the fourfold or twofold changes in singular properties near the lightcone of these global two-point functions across a conjugate plane or a caustic point \cite{CDOW09,BusCas18,HarDri12}. The impulsive plane wave spacetime is an example in which this change of singular properties can be shown in simple terms. We shall elucidate this in the following.

Consider the degenerate case with $\lambda_{1}=\lambda_{2}=-\lambda$. Near the conjugate plane, 
the world function and van Vleck determinant diverge according to Eqs.~(\ref{sigmalimit1}) and (\ref{Deltalimit1}), respectively. Then, the positive Wightman function becomes
\begin{eqnarray}
G_{+}(z,z')&=&-\frac{i}{16\pi^{2}}(u-u_{c})^{-1}(\lambda u_{c}^{2})\int_{0}^{\infty}\frac{dT}{T^{2}}e^{-im^{2}T}e^{\frac{-iu_{c}u'\lambda}{4T(u-u_{c})}\sum_{a=1,2}\left(x^{a}-\frac{u_{c}x'^{a}}{u'}\right)^{2}}\nonumber\\
&&\hskip 70pt e^{\frac{i}{4T}\left\{2\lambda u_{c}u'(v-v')+\sum_{a=1,2}\left[\left(x^{a}-\frac{u_{c}x'^{a}}{u'}\right)^{2}+\lambda^{2}u_{c}^{2}(x'^{a})^{2}\right]\right\}}\label{wightmanlimit1}
\end{eqnarray}
As $u\rightarrow u_{c}$, delta functions emerge in the limit similar to that in Eq.~(\ref{deltalimit}) giving
\begin{eqnarray}
\left.G_{+}(z,z')\right|_{u\rightarrow u_{c}}&=&-\frac{i}{16\pi^{2}}\sqrt{\tilde{\Delta}}\int_{0}^{\infty}\frac{dT}{T}e^{-im^{2}T}e^{i\tilde{\sigma}/2T}\prod_{a=1,2}\delta\left(x^{a}-\frac{u_{c}x'^{a}}{u'}\right)\label{wightmandelta1}
\end{eqnarray}
Instead of the world function $\sigma$ and the van Vleck determinant $\Delta$ which diverge on the conjugate plane, we have the expression $\tilde{\sigma}=-(u_{c}-u')[(v-v')+u_{c}\lambda\vec{x}'^{2}/2u']$ and $\tilde{\Delta}=-(4\pi u_{c}/u')^{2}$ which are both well defined there. Due to the delta functions, the positive Wightman function on the conjugate plane is nonzero only at the point $\vec{x}=u_{c}\vec{x}'/u'$. This can also be interpretated as the focusing effect of the two-point functions on the conjugate plane.

Away from the conjugate plane, the world function and the van Vleck determinant are well-defined. Hence, the expression for the positive Wightman function in Eq.~(\ref{finalWightman}) is also valid even when $z$ is beyond the normal neigborhood of $z'$. In this form it is possible to explore the singular properties of the positive Wightman function near the lightcone in which $\sigma\rightarrow 0$ on both sides of the conjugate plane. To continue, we shall consider separately the real and the imaginary parts of $G_{+}(z,z')$. The real part is related to the Hadamard elementary function. The imaginary part is related to the Pauli-Jordan function which is also used to construct the retarded and the advanced Green functions. 

For the degenerate case that we are considering, $\sqrt{\Delta(u,u')}=(u-u')/(u-u'+uu'\lambda)$. First, this function is real. Moreover, $\sqrt{\Delta(u,u')}>0$ for $u_{c}>u>0$ and $\sqrt{\Delta(u,u')}<0$ for $u>u_{c}$. From Eq.~(\ref{finalWightman}), for $u_{c}>u$, the leading and subleading singularities of the real part of $G_{+}(z,z')$ near the lightcone are
\begin{eqnarray}
{\rm Re}(G_{+}(z,z'))=\frac{\sqrt{\Delta}}{8\pi^{2}\sigma}+\frac{m^{2}}{16\pi^{2}}\sqrt{\Delta}\ {\rm ln}\left(\frac{m^{2}|\sigma|}{2}\right)+\cdots\label{realWightman}
\end{eqnarray}
where $\cdots$ represents the non-singular part.
Hence, across the conjugate plane the leading singularity goes from $1/\sigma\rightarrow-1/\sigma$ and the subleading singularity goes from ln$|\sigma|\rightarrow-$ln$|\sigma|$. In addition, the singular structure of the imaginary part goes like
\begin{eqnarray}
{\rm Im}(G_{+}(z,z'))=-\frac{1}{8\pi}\sqrt{\Delta}\ \delta(\sigma)+\frac{m^{2}}{16\pi}\sqrt{\Delta}\ \theta(-\sigma)+\dots\label{imaginaryWightman}
\end{eqnarray}
Hence, across the conjugate plane the leading singularity goes from $-\delta(\sigma)\rightarrow\delta(\sigma)$ and the subleading singularity goes from $\theta(-\sigma)\rightarrow-\theta(\sigma)$. This represents a twofold singularity structure discussed recently for the retarded Green function in spacetimes with caustics \cite{BusCas18,HarDri12}. 

Next, we examine the nondegenerate case with $\lambda_{1}=-\lambda=-\lambda_{2}$. Here, the divergences of the world function and the van Vleck determinant as $u\rightarrow u_{c}$ are given by Eqs.~(\ref{sigmalimit2}) and (\ref{Deltalimit2}). Then, on the conjugate plane $u=u_{c}$ the positive Wightman function can be expressed as
\begin{eqnarray}
\left.G_{+}(z,z')\right|_{u\rightarrow u_{c}}&=&-\frac{i}{16\pi^{2}}\sqrt{\tilde{\tilde{\Delta}}}\int_{0}^{\infty}\frac{dT}{T^{3/2}}e^{-im^{2}T}e^{i\tilde{\tilde{\sigma}}/2T}\ \delta\left(x^{1}-\frac{u_{c}x'^{1}}{u'}\right)\label{wightmandelta2}
\end{eqnarray}
where $\tilde{\tilde{\Delta}}=i2\pi u_{c} / u'$ and 
\begin{eqnarray}
\tilde{\tilde{\sigma}}&=&-(u_{c}-u')\bigg\{(v-v')+\frac{u_c\lambda}{2u'}(x'^{1})^{2}\nonumber\\
&&\hskip 50pt+\frac{1}{4u_{c}u'\lambda}\left[(1-u'\lambda)(x^{2})^{2}+(1+u_{c}\lambda)(x'^{2})^{2}-2x^{2}x'^{2}
\right]
\bigg\}\label{nondegsigma}
\end{eqnarray}
We can see that on the conjugate plane the positive Wightman function is nonzero along the line enforced by the delta function. This is the focusing effect of the two-point function in this nondegenerate case.

As we have discussed before, the expression for the positive Wightman function in Eq.~(\ref{finalWightman}) is valid everywhere except on the conjugate plane. However, in this nondegenerate case, $\sqrt{\Delta(u,u')}=(u-u')(u-u'+uu'\lambda)^{-1/2}(u-u'-uu'\lambda)^{-1/2}$ which is real and positive for $u_{c}>u>0$ but after crossing the conjugate plane it is multiplied by $-i$ due to the factor $(u-u_{c})^{-1/2}$. In this way, the real and the imaginary part of the positive Wightman function will interchange. Before crossing the conjugate plane, the real part of $G_{+}(z,z')$ is given by Eq.~(\ref{realWightman}). After crossing the plane, we have
\begin{eqnarray}
{\rm Re}(G_{+}(z,z'))=-\frac{1}{8\pi}\sqrt{|\Delta|}\ \delta(\sigma)+\frac{m^{2}}{16\pi}\sqrt{|\Delta|}\ \theta(-\sigma)+\dots
\end{eqnarray}
Hence, across the conjugate plane in this nondegenerate case, the leading singularity goes from $1/\sigma\rightarrow-\delta(\sigma)$, while the subleading singularity goes from ${\rm ln}|\sigma|\rightarrow\theta(-\sigma)$.
Similarly, for the imaginary part, after crossing the conjugate plane
\begin{eqnarray}
{\rm Im}(G_{+}(z,z'))=-\frac{\sqrt{|\Delta|}}{8\pi^{2}\sigma}-\frac{m^{2}}{16\pi^{2}}\sqrt{|\Delta|}\ {\rm ln}\left(\frac{m^{2}|\sigma|}{2}\right)+\cdots
\end{eqnarray}
The leading singularity goes from $-\delta(\sigma)\rightarrow -1/\sigma$ and the subleading singularity from $\theta(-\sigma)\rightarrow -{\rm ln}|\sigma|$. This would constitute a fourfold transformation across the conjugate plane: $\delta(\sigma)\rightarrow 1/\sigma\rightarrow-\delta(\sigma)\rightarrow-1/\sigma\rightarrow\delta(\sigma)\rightarrow\cdots$ for the leading singularity and $\theta(-\sigma)\rightarrow-{\rm ln}|\sigma|\rightarrow-\theta(-\sigma)\rightarrow{\rm ln}|\sigma|\rightarrow\theta(-\sigma)\rightarrow\cdots$ for the subleading singularity. This is related to the recently found fourfold transformation of the retarded Green function across caustics in black hole spacetimes \cite{CDOW09}.

Therefore, we see that twofold or fourfold transformations of singularities near the lightcone of the two-point functions across conjugate planes are mainly controlled by the behavior of the van Vleck determinant \cite{BusCas18}. Our consideration of impulsive plane wave spacetimes gives simple examples of this phenomenon. Moreover, through the Penrose limiting procedure, it has been argued that the singularity structure elaborated above is actually a property for two-point functions, especially the retarded Green function, in general curved spacetimes \cite{HarDri12}.


\section{Noise kernel: Across the conjugate plane}\label{noise}

In the semiclassical gravity, the effect of quantum fields on the classical spacetime is given by the expectation value of the stress energy tensor $T_{\mu\nu}$ \cite{BirDav82}. This expectation value is usually divergent. Regularization and renormalization procedures have to be devised to obtain finite results. Using the point-separation method \cite{Christensen76}, one can define the renormalized expectation value of the stress energy tensor as
\begin{eqnarray}
\langle T_{\mu\nu}(z)\rangle_{\rm ren}=\lim_{z'\rightarrow z}\frac{1}{2}{\cal T}_{\mu\nu}G^{(1)}_{\rm ren}(z,z')
\end{eqnarray}
where, for the minimally coupled scalar field that we are considering, the differential operator is given by
\begin{eqnarray}
{\cal T}_{\mu\nu}=\frac{1}{2}(g_{\mu}^{\ \alpha'}\nabla_{\alpha'}\nabla_{\nu}+g_{\nu}^{\ \alpha'}\nabla_{\alpha'}\nabla_{\mu})-\frac{1}{2}g_{\mu\nu}g^{\rho\alpha'}\nabla_{\rho}\nabla_{\alpha'}-\frac{1}{2}m^{2}g_{\mu\nu}
\end{eqnarray}
where $g_{\mu}^{\ \alpha'}(z,z')$ is the parallel propagator. $G^{(1)}_{\rm ren}(z,z')$ is the renormalized Hadamard elementary function. Note that $G^{(1)}(z,z')$ is twice the real part of the positive Wightman function. $G^{(1)}_{\rm ren}(z,z')$ is obtained by subtracting the divergent part of $G^{(1)}(z,z')$ when the $z'\rightarrow z$ is taken. Under the assumption that $\langle T_{\mu\nu}\rangle_{\rm ren}$ should vanish for a free field in the Minkowski vacuum, we should have $G^{(1)}_{\rm ren}=0$ in Minkowski spacetime. From Section \ref{wightman}, we have found that the positive Wightman functions for $0>u>u'$ and $u>u'>0$ are just the Minkowski one. Therefore, we have $\langle T_{\mu\nu}\rangle_{\rm ren}=0$ in the impulsive plane wave spacetime.


Even though the renormalized expectation value of the stress energy tensor is zero in the impulsive plane wave spacetime, it does not mean that its fluctuations and correlations are also vanishing. It is therefore interesting to examine the properties of the stress energy tensor correlators. In semicalssical gravity, only the mean value of $T_{\mu\nu}$ is taken into account. To further consider  the effects of fluctuations and correlations, the theory of stochastic gravity is put forth by Hu and Verdaguer \cite{HuVer08,HuVer20}. This is an open quantum system approach in which gravity is the system and quantum field is the environment. The influence of the environment on the system is manifested as noise and dissipative effects. The corresponding noise kernel is actually given by the correlator of the stress energy tensor $N_{\mu\nu\alpha'\beta'}(z,z')=\langle( T_{\mu\nu}(z)-\langle T_{\mu\nu}(z)\rangle)(T_{\alpha'\beta'}(z')-\langle T_{\alpha'\beta'}(z')\rangle)\rangle$. In the following we shall examine the properties of this noise kernel in the impulsive plane wave spacetime.

\subsection{Noise kernel by point-separation method}
Using the point-separation method, a general formula for the noise kernel is given in \cite{PhiHu01}. For a minimally coupled scalar field, it is
\begin{eqnarray}
N_{\mu\nu\alpha'\beta'}={\rm Re}\left(\tilde{N}_{\mu\nu\alpha'\beta'}+g_{\mu\nu}\tilde{N}_{\alpha'\beta'}+g_{\alpha'\beta'}\tilde{N}_{\mu\nu}'+g_{\mu\nu}g_{\alpha'\beta'}\tilde{N}\right)\label{noisekernel1}
\end{eqnarray}
where
\begin{eqnarray}
\tilde{N}_{\mu\nu\alpha'\beta'}&=&G_{;\mu\alpha'}G_{;\nu\beta'}+G_{;\mu\beta'}G_{;\nu\alpha'}\\
\tilde{N}'_{\mu\nu}&=&-G_{;\mu\alpha'}G_{;\nu}^{\ \ \alpha'}-m^{2}G_{;\mu}G_{;\nu}\\
\tilde{N}&=&\frac{1}{2}G_{;\mu\alpha'}G^{\,;\mu\alpha'}+\frac{1}{2}m^{2}\left(G_{;\mu}G^{\,;\mu}+G_{;\alpha'}G^{\,;\alpha'}\right)+\frac{1}{2}m^{4}G^{2}\label{noisekernel2}
\end{eqnarray}
where $G$ is the positive Wightman function. Note that the $+$ index has been omitted to simplify notation.

Here in our case of the impulsive plane wave spacetime, the positive Wightman function as given in Eq.~(\ref{finalWightman}), can be written as $G(z,z')=\sqrt{\Delta(u,u')}\,g(\sigma)$ with the van Vleck determinant a function of $u$ and $u'$ and the rest $g(\sigma)$ a function of the world function $\sigma$. Since $G$ satisfies the Klein-Gordon equation, we have $2\sigma g''+4g'-m^{2}g=0$.
Together with the identities: $\sigma_{,\mu}\sigma^{\ ,\mu}=2\sigma$ and $\sigma_{,\mu}({\rm ln}\Delta)^{\, ,\mu}=4-\sigma_{,\mu}^{\ \ \mu}$, the noise kernel components can be expressed as follows.
\begin{eqnarray}
\tilde{N}_{\mu\nu\alpha'\beta'}&=&\Delta\bigg\{\bigg[g'\left(\sigma_{,\mu\alpha'}-\frac{2\sigma_{,\mu}\sigma_{,\alpha'}}{\sigma}+\frac{1}{2}\sigma_{,\mu}({\rm ln}\Delta)_{,\alpha'}+\frac{1}{2}\sigma_{,\alpha'}({\rm ln}\Delta)_{,\mu}\right)\nonumber\\
&&\hskip 30pt +g\left(\frac{m^{2}\sigma_{,\mu}\sigma_{,\alpha'}}{2\sigma}+\frac{1}{2}({\rm ln}\Delta)_{,\mu\alpha'}+\frac{1}{4}({\rm ln}\Delta)_{,\mu}({\rm ln}\Delta)_{,\alpha'}\right)\bigg]\nonumber\\
&&\hskip 20pt \bigg[g'\left(\sigma_{,\nu\beta'}-\frac{2\sigma_{,\nu}\sigma_{,\beta'}}{\sigma}+\frac{1}{2}\sigma_{,\nu}({\rm ln}\Delta)_{,\beta'}+\frac{1}{2}\sigma_{,\beta'}({\rm ln}\Delta)_{,\nu}\right)\nonumber\\
&&\hskip 30pt +g\left(\frac{m^{2}\sigma_{,\nu}\sigma_{,\beta'}}{2\sigma}+\frac{1}{2}({\rm ln}\Delta)_{,\nu\beta'}+\frac{1}{4}({\rm ln}\Delta)_{,\nu}({\rm ln}\Delta)_{,\beta'}\right)\bigg]\nonumber\\
&&\ \ \ +(\mu\leftrightarrow\nu)\bigg\}\label{noisekernel3}
\end{eqnarray}

\begin{eqnarray}
\tilde{N}'_{\mu\nu}&=&\Delta\bigg\{ g'^{2}\bigg[-\frac{1}{2}\sigma_{,\mu\alpha'}\sigma_{,\nu}^{\ \ \alpha'}+\frac{2\sigma_{,\mu}\sigma_{,\nu}}{\sigma}-\frac{1}{2}m^{2}\sigma_{,\mu}\sigma_{,\nu}-\frac{\sigma_{,\mu}\sigma_{,\nu}\sigma_{,\alpha'}^{\ \ \alpha'}}{\sigma}-\frac{1}{2}\sigma_{,\mu\alpha'}\sigma_{,\nu}({\rm ln}\Delta)^{,\alpha'}\nonumber\\
&&\hskip 50pt +\frac{1}{2}\sigma_{,\mu}({\rm ln}\Delta)_{,\nu}+\frac{1}{4}\sigma_{,\mu}\sigma_{,\alpha'}^{\ \ \alpha'}({\rm ln}\Delta)_{,\nu}-\frac{1}{4}\sigma({\rm ln}\Delta)_{,\mu}({\rm ln}\Delta)_{,\nu}\bigg]\nonumber\\
&&\ \ \ +g g'\bigg[\frac{m^{2}\sigma_{,\mu}\sigma_{,\nu}}{2\sigma}+\frac{m^{2}\sigma_{,\mu}\sigma_{,\nu}\sigma_{,\alpha'}^{\ \ \alpha'}}{4\sigma}-\left(m^{2}+\frac{\sigma_{,\alpha'}^{\ \ \alpha'}}{2\sigma}-\frac{2}{\sigma}\right)\sigma_{,\mu}({\rm ln}\Delta)_{,\nu}\nonumber\\
&&\hskip 50pt +\frac{\sigma_{,\mu}\sigma_{,\alpha'}}{\sigma}({\rm ln}\Delta)_{,\nu}^{\ \ \alpha'}-\frac{1}{2}\sigma_{,\mu\alpha'}({\rm ln}\Delta)_{,\nu}^{\ \ \alpha'}-\frac{1}{2}\left(1-\frac{1}{4}\sigma_{,\alpha'}^{\ \ \alpha'}\right)({\rm ln}\Delta)_{,\mu}({\rm ln}\Delta)_{,\nu}\nonumber\\
&&\hskip 50pt -\frac{1}{4}\sigma_{,\mu\alpha'}({\rm ln}\Delta)_{,\nu}({\rm ln}\Delta)^{,\alpha'}-\frac{1}{4}\sigma_{,\alpha'}({\rm ln}\Delta)_{,\mu}({\rm ln}\Delta)_{,\nu}^{\ \ \alpha'}\bigg]\nonumber\\
&&\ \ \ +g^{2}\bigg[-\frac{m^{4}\sigma_{,\mu}\sigma_{,\nu}}{4\sigma}-\frac{m^{2}\sigma_{,\mu}}{2\sigma}({\rm ln}\Delta)_{,\nu}+\frac{m^{2}\sigma_{,\mu}\sigma_{,\alpha'}^{\ \ \alpha'}}{8\sigma}({\rm ln}\Delta)_{,\nu}
-\frac{m^{2}\sigma_{,\mu}\sigma_{,\alpha'}}{4\sigma}({\rm ln}\Delta)_{,\nu}^{\ \ \alpha'}\nonumber\\
&&\hskip 50pt 
-\frac{1}{8}m^{2}({\rm ln}\Delta)_{,\mu}({\rm ln}\Delta)_{,\nu}\bigg]+(\mu\leftrightarrow\nu)\bigg\}\label{noisekernel4}
\end{eqnarray}

\begin{eqnarray}
\tilde{N}&=&\frac{\Delta}{2}\bigg\{
g'^{2}\bigg[-8+\sigma_{,\mu}^{\ \ \mu}+\sigma_{,\alpha'}^{\ \ \alpha'}+\frac{1}{2}\sigma_{,\mu}^{\ \ \mu}\sigma_{,\alpha'}^{\ \ \alpha'}+\sigma_{,\mu\alpha'}\sigma^{,\mu\alpha'}+4m^2\sigma\bigg]\nonumber\\
&&\ \ \ +g g'\bigg[10m^{2}-2m^{2}\sigma_{,\mu}^{\ \ ,\mu}-2m^{2}\sigma_{,\alpha'}^{\ \ \alpha'}-\frac{16}{\sigma}+\frac{4\sigma_{,\mu}^{\ \ \mu}}{\sigma}+\frac{4\sigma_{,\alpha'}^{\ \ \alpha'}}{\sigma}-\frac{\sigma_{,\mu}^{\ \ \mu}\sigma_{,\alpha'}^{\ \ \alpha'}}{\sigma}\nonumber\\
&&\hskip 50pt -\frac{2\sigma_{,\mu}\sigma_{,\alpha'}}{\sigma}({\rm ln}\Delta)^{,\mu\alpha'}+\sigma_{,\mu\alpha'}({\rm ln}\Delta)^{,\mu\alpha'}+\frac{1}{2}\sigma_{,\mu\alpha'}({\rm ln}\Delta)^{,\mu}({\rm ln}\Delta)^{,\alpha'}\bigg]
\nonumber\\
&&\ \ \ +g^{2}\bigg[2m^{4}+\frac{4m^{2}}{\sigma}-\frac{m^{2}\sigma_{,\mu}^{\ \ \mu}}{\sigma}-\frac{m^{2}\sigma_{,\alpha'}^{\ \ \alpha'}}{\sigma}+\frac{m^{2}\sigma_{,\mu}^{\ \ \mu}\sigma_{,\alpha'}^{\ \ \alpha'}}{4\sigma}+\frac{m^{2}\sigma_{,\mu}\sigma_{,\alpha'}}{2\sigma}({\rm ln}\Delta)^{,\mu\alpha'}\bigg]\nonumber\\ \label{noisekernel5}
\end{eqnarray}
Since we are interested in the correlation function across the shock, we take $u'<0$ and $u>0$. In the next subsection, we shall consider the singularity of the noise kernel near the lightcone. Here, we concentrate on the case with $\sigma\neq 0$, and 
\begin{eqnarray}
g(\sigma)=\frac{im}{8\pi\sqrt{-2\sigma}}\theta(-\sigma)H_{1}^{(2)}(m\sqrt{-2\sigma})+\frac{m}{4\pi^{2}\sqrt{2\sigma}}\theta(\sigma)K_{1}(m\sqrt{2\sigma})
\end{eqnarray}
As we can see from Eqs.~(\ref{noisekernel1}) to (\ref{noisekernel5}), the expressions for the noise kernel components are in general quite complicated. We therefore look at some limiting cases below for small and large geodesic separations. These would apply to cases in and beyond the normal neighborhood.

For small geodesic distance, or $\sigma\ll 1$, the function 
\begin{eqnarray}
g(\sigma)=\frac{1}{8\pi^{2}\sigma}+\cdots\Rightarrow g'(\sigma)=-\frac{1}{8\pi^{2}\sigma^{2}}+\cdots
\end{eqnarray}
This indicates that near the lightcone the behavior is like that of a massless scalar field in Minkowski spacetime, except that the van Vleck determinant is not of unity and the world function is given by Eq.~(\ref{finalsigma}) instead. As a result the leading contribution to the noise kernel is given by terms proportional to $g'^{2}$ and with the higher power of $\sigma$ in the denominator. 
\begin{eqnarray}
N_{\mu\nu\alpha'\beta'}&=&\frac{\Delta}{16\pi^{4}}\bigg(
\frac{\sigma_{,\mu}\sigma_{,\nu}\sigma_{,\alpha'}\sigma_{,\beta'}}{\sigma^{6}}\bigg)
+\cdots\label{smallsigma}
\end{eqnarray}
To give an explicit expression, we consider the energy density correlator $N_{uuu'u'}=\langle T_{uu}T_{u'u'}\rangle$. To make the expression manageable, we set, without loss of generality, $z'=(u',0,\vec{0})$. Further simplification can be obtained if we set $\lambda_{1}=\lambda_{2}=-\lambda$, that is, for the degenerate case. To approach the lightcone, we take
\begin{eqnarray}
v=\frac{(1+u'\lambda)}{2(u-u'+uu'\lambda)}\sum_{a}(x^{a})^{2}+\delta v
\end{eqnarray}
where $\delta v\rightarrow 0$ corresponds to the small $\sigma$ limit. Then, we have
\begin{eqnarray}
N_{uuu'u'}&=&\left(\frac{(u-u')^{6}(1+u'\lambda)^{4}(\sum_{a}(x^{a})^{2})^{4}}{256\pi^{4}(u-u'+uu'\lambda)^{10}}\right)\frac{1}{\sigma^{6}}\nonumber\\
&&\ \ +\frac{(u-u')^{5}(1+u'\lambda)^{3}(\sum_{a}(x^{a})^{2})^{3}}{512\pi^{4}(u-u'+uu'\lambda)^{10}}\bigg[4(u-u'+uu'\lambda)(-4+\lambda(u-u'+uu'\lambda))\nonumber\\
&&\hskip 100pt -m^{2}(u-u')(1+u'\lambda)\sum_{a}(x^{a})^{2}\bigg]\frac{1}{\sigma^{5}}+\cdots
\end{eqnarray}
which is consistent with Eq.~(\ref{smallsigma}).

For large timelike geodesic separation, $\sigma\rightarrow -\infty$, 
\begin{eqnarray}
g(\sigma)&=&\frac{im}{8\pi\sqrt{-2\sigma}}H^{(2)}_{1}(m\sqrt{-2\sigma})\nonumber\\
&=&e^{-im\sqrt{-2\sigma}-i3\pi/4}\left(-\frac{3i}{64 (2)^{3/4}\pi^{3/2}m^{1/2}}\right)\frac{1}{(-\sigma)^{5/4}}\left(1+\frac{i8(2)^{1/2}m}{3}(-\sigma)^{1/2}+\cdots\right)\nonumber\\
\end{eqnarray}
Therefore, the noise kernel components $N_{\mu\nu\alpha'\beta'}$ are dominated by $\sin(2m\sqrt{-2\sigma})$ or $\cos(2m\sqrt{-2\sigma})$ with negative powers of $(-\sigma)$. For example, for the $N_{uuu'u'}$, as $\sigma\rightarrow-\infty$,
\begin{eqnarray}
N_{uuu'u'}&=&\frac{m^{3}}{512\pi^{3}(1+u'\lambda)^{4}}\frac{1}{(-\sigma)^{3/2}}\bigg[\sqrt{2}\Big(m^{2}(1+2u'\lambda)-\lambda^{2}(2-m^{2}u'^{2})\sin(2m\sqrt{-2\sigma}\Big)\nonumber\\
&&\hskip 140pt -4m\lambda(1+u'\lambda)\cos(2m\sqrt{-2\sigma})\bigg]+\cdots
\end{eqnarray}
Here, we have assumed that $z'=(u',0,\vec{0})$ and $z=(u,v,\vec{0})$ to simplify the expression. Then, $\sigma=-(u-u')v$ and the limit $\sigma\rightarrow -\infty$ is reached by taking $(u-u')= (-\sigma)^{1/2}\rightarrow\infty$ and $v=(-\sigma)^{1/2}\rightarrow\infty$.

For large spacelike geodesic separation, $\sigma\rightarrow\infty$,
\begin{eqnarray}
g(\sigma)&=&\frac{m}{4\pi^{2}\sqrt{2\sigma}}K_{1}(m\sqrt{2\sigma})\nonumber\\
&=&\frac{\pi^{1/2}}{2^{3/4}m^{1/2}}\frac{e^{-m\sqrt{2\sigma}}}{\sigma^{1/4}}\bigg(1+\frac{3\sqrt{2}}{16m}\frac{1}{\sqrt{\sigma}}+\cdots\bigg)\label{largespace}
\end{eqnarray}
The noise kernel in this limit is given mainly by $e^{-2m\sqrt{2\sigma}}$ with negative powers of $\sigma$. Again, we look at the energy density correlator with $(u-u')=(\sigma)^{1/2}\rightarrow\infty$ and $v=(\sigma)^{1/2}\rightarrow\infty$. The leading contribution is
\begin{eqnarray}
N_{uuu'u'}=\frac{m^{3}}{256\sqrt{2}\pi^{3}(1+u'\lambda)^{4}}\Big(\sqrt{2}\lambda-m(1+u'\lambda)\Big)^{2}\left(\frac{e^{-2m\sqrt{2\sigma}}}{\sigma^{3/2}}\right)+\cdots
\end{eqnarray}
which is indeed suppressed by a factor $e^{-2m\sqrt{2\sigma}}$.

\subsection{On and across the conjugate plane}
From what we have discussed in Sections II and III, both $\sigma(z,z')$ and $\Delta(u,u')$ diverge as one approaches the conjugate plane at $u=u_{c}$. However, the positive Wightman function can still be described on the conjugate plane in terms of Dirac delta functions located at specific points or lines on the $x^{1}$-$x^{2}$ plane. Since the noise kernel $N_{\mu\nu\alpha'\beta'}(z,z')$ are expressed in terms of various derivatives of the Wightman functions, we expect that on the conjugate plane $N_{\mu\nu\alpha'\beta'}(z,z')$ could also be expressed in terms of product of delta functions and its derivatives.

To make the discussion more concrete, we consider the degenerate case with $\lambda_{1}=\lambda_{2}=-\lambda$. Near the conjugate plane with $u=u_{c}=u'/(1+u'\lambda)$, the world function diverges as $(u-u_{c})^{-1}$ as shown in Eq.~(\ref{sigmalimit1}). To study the noise kernel near and on the conjugate plane, it is best to use the representation of the positive Wightman function in Eqs.~(\ref{wightmanlimit1}) and (\ref{wightmandelta1}). From Eqs.~(\ref{noisekernel1}) and (\ref{noisekernel2}), we can see that the noise kernel is basically given by the derivatives of the Wightman function. 
We take the noise kernel component $N_{uuu'u'}$ as an example. Using Eq.~(\ref{wightmanlimit1}),
\begin{eqnarray}
N_{uuu'u'}&=&{\rm Re}(2G_{,uu'}^2)\nonumber\\
&=&{\rm Re}\Bigg[2\bigg(\frac{i\lambda u_{c}^{4}}{8\pi^{2}u'^{2}}(u-u_{c})^{-3}\int_{0}^{\infty}\frac{dT}{T^{2}}e^{-im^{2}T}e^{\frac{-iu_{c}u'\lambda}{4T(u-u_{c})}\sum_{a=1,2}\left(x^{a}-\frac{u_{c}x'^{a}}{u'}\right)^{2}}\nonumber\\
&&\hskip 70pt e^{\frac{i}{4T}\left\{2\lambda u_{c}u'(v-v')+\sum_{a=1,2}\left[\left(x^{a}-\frac{u_{c}x'^{a}}{u'}\right)^{2}+\lambda^{2}u_{c}^{2}(x'^{a})^{2}\right]\right\}}\bigg)^{2}+\cdots\Bigg]
\end{eqnarray}
where we have shown the most divergent term as $u\rightarrow u_{c}$. Note that the divergent quantity here can be expressed as derivatives on delta function.
\begin{eqnarray}
&&\lim_{u\rightarrow u_{c}}(u-u_{c})^{-3}\ e^{\frac{-iu_{c}u'\lambda}{4T(u-u_{c})}\sum_{a=1,2}\left(x^{a}-\frac{u_{c}x'^{a}}{u'}\right)^{2}}\nonumber\\
&=&\left(-\frac{T^{2}}{2u_{c}^{2}u'^{2}\lambda^{2}}\right)\vec{\nabla}^{2}\vec{\nabla}^{2}\lim_{u\rightarrow u_{c}}(u-u_{c})^{-1}\ e^{\frac{-iu_{c}u'\lambda}{4T(u-u_{c})}\sum_{a=1,2}\left(x^{a}-\frac{u_{c}x'^{a}}{u'}\right)^{2}}\nonumber\\
&=&\left(\frac{i2\pi T^{3}}{u_{c}^{3}u'^{3}\lambda^{3}}\right)\vec{\nabla}^{2}\vec{\nabla}^{2}\delta(\vec{x}-\frac{u_{c}\vec{x}'}{u'})
\end{eqnarray}
where $\vec{\nabla}^{2}=\partial^{2}/\partial(x^{1})^{2}+\partial^{2}/\partial(x^{2})^{2}$. As a result, the noise kernel component is given by
\begin{eqnarray}
N_{uuu'u'}&=&{\rm Re}\Bigg\{2\bigg[\left(\frac{i}{16\pi^{2}\lambda^{2} u'^{4}}\right) \vec{\nabla}^{2}\vec{\nabla}^{2}\delta\left(\vec{x}-\frac{u_{c}\vec{x}'}{u'}\right)\sqrt{\tilde{\Delta}}\int_{0}^{\infty}dT\, T\,e^{-im^{2}T}e^{\frac{i}{2T}\tilde{\sigma}}\bigg]^{2}+\cdots\Bigg\}\nonumber\\
\end{eqnarray}
where $\tilde{\Delta}$ and $\tilde{\sigma}$ are defined as in Eq.~(\ref{wightmandelta1}). This noise kernel component is indeed expressed in terms of delta function and its derivatives, and we expect the other component to behave similarly.

Next, we study the singular structure of the noise kernel near the lightcone. For the Wightman function, this is given by Eqs.~(\ref{realWightman}) and (\ref{imaginaryWightman}). From Eqs.~(\ref{noisekernel1}) and (\ref{noisekernel2}), we can see that the noise kernel is basically the square of the second derivative of the Wightman function. Hence, the most divergent contribution to the noise kernel near the lightcone would be $\sim\Delta[\delta''(\sigma)]^{2}$, while the next subleading term would be $\sim\Delta[\delta''(\sigma)][\delta'(\sigma)]$. For the degenerate case ($\lambda_{1}=\lambda_{2}=-\lambda$) that we are working on here, as evident from Eq.~(\ref{Deltalimit1}), the van Vleck determinant does not change sign crossing the conjugate plane. Therefore, the singularity structure of the noise kernel near the lightcone are the same on both sides of the plane.

In the nondegenerate case, $\lambda_{1}=-\lambda=-\lambda_{2}$, the Wightman function near and on the conjugate plane $u=u_{c}$ are given by Eqs.~(\ref{wightmandelta2}) and (\ref{nondegsigma}). Using a similar consideration as that for the degenerate case above, the noise kernel component $N_{uuu'u'}$ on the conjugate plane can be expressed as
\begin{eqnarray}
&&N_{uuu'u'}\nonumber\\
&=&{\rm Re}\bigg\{2\bigg[\left(-\frac{i}{16\pi^{2}\lambda^{2}u'^{4}}\right)\left(\frac{\partial}{\partial x^{1}}\right)^{4}\delta\left(x^{1}-\frac{u_{c}x'^{1}}{u'}\right)\sqrt{\tilde{\tilde{\Delta}}}\int_{0}^{\infty}dT T^{1/2}e^{-im^{2}T}e^{\frac{i}{2T}\tilde{\tilde{\sigma}}}\bigg]^{2}+\cdots\bigg\}\nonumber\\
\end{eqnarray}
where we have shown the most divergent term in the noise kernel component. The functions $\tilde{\tilde{\Delta}}$ and $\tilde{\tilde{\sigma}}$ are defined as in Eq.~(\ref{wightmandelta2}).

For the singular structure of the noise kernel near the lightcone in the nondegenerate case, we see from Eq.~(\ref{Deltalimit2}) that the van Vleck determinant changes sign after crossing the conjugate plane. Hence, for $u<u_{c}$, the leading singularity of the noise kernel near the lightcone is $|\Delta|[\delta''(\sigma)]^{2}$ and the subleading one is $|\Delta|[\delta''(\sigma)][\delta'(\sigma)]$. Whereas for $u>u_{c}$, the leading singularity becomes $-|\Delta|[\delta''(\sigma)]^{2}$ and the subleading one $-|\Delta|[\delta''(\sigma)][\delta'(\sigma)]$.


\section{Conclusions and discussions}
The main results in this paper are about the properties of the Wightman function and the stochastic gravity noise kernel of a scalar field in impulsive plane wave spacetimes. The existence of conjugate planes where caustic points are located is an important feature of these spacetimes. We have shown that on these conjugate planes the two-point correlation functions, in particular the positive Wightman function, are described by Dirac delta functions. This means that the Wightman function vanishes on the conjugate plane except at the focusing points or lines where the function diverges. On the other hand, the singularity structure of the Wightman function near the lightcone can be traced to the behaviors of the van Vleck determinant. In the simple setting of an impulsive plane wave spacetime, the twofold or fourfold transformations of the singularity structure can be shown quite explicitly. Through the Penrose limit, one can argue that these structures are also valid in general curved spacetimes.

In the theory of stochastic gravity, fluctuation and correlation effects of the quantum field are accounted for by the dissipation and the noise kernels. The noise kernel is given by the correlation function of the quantum field stress energy tensor. In this paper, we have worked out the explicit form of this noise kernel in terms of the world function $\sigma(z,z')$ and the van Vleck determinant $\Delta(u,u')$ as well as their derivatives. The leading contribution of the noise kernel is of $1/\sigma^{6}$ for small $\sigma$. For large spacelike $\sigma$, the noise kernel is suppressed by $e^{-2m\sqrt{2\sigma}}$, while for large timelike $\sigma$, its leading contributions are proportional to $(-\sigma)^{-3/2}$ multiplying $\sin(2m\sqrt{-2\sigma})$ or $\cos(2m\sqrt{-2\sigma})$. On the conjugate plane, the leading term of the noise kernel is given by the square of the fourth derivative of the delta function. For the singularity structure near the lightcone in the degenerate case, the leading singularity of the noise kernel is $\sim[\delta''(\sigma)]^{2}$ on both sides of the conjugate plane since the van Vleck determinant $\Delta(u,u')$ does not change sign crossing the conjugate plane in this case. On the other hand, in the nondegenerate case, $\Delta(u,u')$ changes sign when crossing the conjugate plane. Hence, in this case, the singularity changes from $\sim[\delta''(\sigma)]^{2}$ to $\sim-[\delta''(\sigma)]^{2}$. These divergences can be dealt with if we consider the noise kernel as a bidistribution, either by point-separation method \cite{Christensen76,PhiHu01} or integrating with smearing functions \cite{HuRou07}.

Under the Penrose limit, the spacetime near a null geodesic can be viewed as a plane wave spacetime. Hence, we like to apply the semiclassical stochastic gravity theory to plane wave spacetime to capture the essential physics, like for example, the quantum energy inequalities \cite{ForRom95}, of a quantum field near a null geodesic. Our study of the stochastic gravity noise kernel can be viewed as a first step towards this goal. In our future works, we plan to further our considerations to include the dissipation kernel which is related to the retarded Green function, and also to the influence action which summaries the effects of the quantum field on the classical spacetime. Furthermore, we would extend our study to more general plane wave spacetimes, for example, that of sandwich waves in which the wave profiles are of finite extend \cite{Gibbons75,GarVer91}. We are 
also interested in spacetimes which are Penrose limit near spacetime singularities \cite{Blau11}. For example, the singular homogeneous plane wave spacetime is the Penrose limit near a black hole singularity \cite{BBOP04}. This would be relevant to our future investigations of the quantum field effects near spacetime singularities.





%
%
\acknowledgments
\noindent{The author would like to thank Bei-Lok Hu for useful discussions. This work is supported in part by the National Council of Science and
Technology, Taiwan, ROC, under the Grants NCST110-2112-M-032-009 and
NCST112-2112-M-032-006.}

%
%


\end{document}